\documentstyle[preprint,revtex,eqsecnum]{aps}
\tightenlines
\begin{document}
\draft

\preprint{SLAC--PUB--5897}
\medskip
\preprint{September 1992}
\medskip
\preprint{T/E}

\begin{title}
Second Order Power Corrections in the Heavy Quark\\
Effective Theory\\
I. Formalism and Meson Form Factors
\end{title}

\author{Adam F. Falk and Matthias Neubert}
\begin{instit}
Stanford Linear Accelerator Center\\
Stanford University, Stanford, California 94309
\end{instit}

\begin{abstract}

In the heavy quark effective theory, hadronic matrix elements of
currents between two hadrons containing a heavy quark are expanded in
inverse powers of the heavy quark masses, with coefficients that are
functions of the kinematic variable $v\cdot v'$. For the ground state
pseudoscalar and vector mesons, this expansion is constructed at order
$1/m_Q^2$. A minimal set of universal form factors is defined in terms
of matrix elements of higher dimension operators in the effective
theory. The zero recoil normalization conditions following from vector
current conservation are derived. Several phenomenological applications
of the general results are discussed in detail. It is argued that at
zero recoil the semileptonic decay rates for $B\to D\,\ell\,\nu$ and
$B\to D^*\ell\,\nu$ receive only small second order corrections, which
are unlikely to exceed the level of a few percent. This supports the
usefulness of the heavy quark expansion for a reliable determination of
$V_{cb}$.

\end{abstract}

\centerline{(Submitted to Physical Review D)}
\newpage
\narrowtext

\section{Introduction}

Recent developments in the theory of heavy quarks have increased the
prospects both for a reliable determination of some of the fundamental
parameters of the standard model, and for a study of nonperturbative
QCD in the weak decays of heavy mesons and baryons. The excitement is
driven by the discovery of a spin-flavor symmetry for heavy quarks that
QCD reveals in the limit where the quark mass $m_Q\to\infty$, in which
certain properties of a hadron containing the heavy quark become
independent of its mass and spin \cite{Volo,Isgu}. These symmetries are
responsible for restrictive relations among weak decay amplitudes and
reduce the number of independent form factors. The description of
semileptonic transitions between two ground state heavy mesons
\cite{Isgu,Falk} or baryons \cite{IWbar,Gebar,MRR,Main} becomes
particularly simple. In the limit where the heavy quark masses are much
larger than any other hadronic scale in the process, the large set of
hadronic form factors is reduced to a single universal function of the
kinematic variable $v\cdot v'$, which measures the change of velocities
that the heavy hadrons undergo during the transition. It depends on the
quantum numbers of the light degrees of freedom, but not on the heavy
quark masses and spins. In addition, the conservation of the vector
current implies that this celebrated Isgur-Wise form factor is
normalized at zero recoil, allowing model-independent predictions
unaffected by hadronic uncertainties.

Clearly, a careful analysis of at least the leading symmetry-breaking
corrections is essential for any phenomenological application of the
heavy quark symmetries. An elegant framework in which to analyze such
corrections is provided by the so-called heavy quark effective theory,
which allows for a systematic expansion of decay amplitudes in powers
of $1/m_Q$ \cite{Eich,Lepa,Grin,Geor,Korn,Mann,FGL}. The coefficients
in this expansion are given by matrix elements of operators in the
effective theory and can be parameterized in terms of universal form
factors, which characterize the properties of the light degrees of
freedom in the background of the static color source provided by the
heavy quark. At leading order one recovers the Isgur-Wise limit, in
which only a single function remains. But already at order $1/m_Q$ one
encounters a larger set of universal form factors, which affect all but
very few of the symmetry predictions that hold in the infinite quark
mass limit \cite{Luke,GGW,SR2}. An understanding of these functions is
at the heart of nonperturbative QCD, but it is ultimately necessary for
any quantitative analysis based on heavy quark symmetries. In the
future, one might hope to compute the universal form factors from first
principles by using a formulation of the effective theory on a lattice
\cite{Eich,Allt,Alex}. In the meantime, QCD sum rules \cite{SVZ} offer
a less ambitious approach to this problem, and have recently been
employed to study the decay constants of heavy mesons, the Isgur-Wise
form factor, and the universal functions that appear at order $1/m_Q$
in the heavy quark expansion \cite{SR2,SR,BBBD,Buch,Rady,Sublea}. One
can also gain valuable information about symmetry-breaking corrections
from measurements of certain ratios of form factors \cite{Sublea}.

In this paper we analyze current-induced transitions between ground
state heavy mesons at order $1/m_Q^2$ in the heavy quark expansion.
Such an analysis is particularly relevant for the important cases where
the leading $1/m_Q$ corrections are known to vanish at zero recoil. An
example is the semileptonic decay $B\to D^*\ell\,\nu$, which therefore
seems ideal for a measurement of the weak mixing parameter $V_{cb}$
\cite{Volo,MN}. In Sec.~\ref{sec:2} we discuss the parameters of the
effective theory that appear at subleading order. The general method of
computing power corrections is outlined in Sec.~\ref{sec:3}, together
with a review of the analysis of the $1/m_Q$ corrections to transitions
between heavy mesons. In Sec.~\ref{sec:4} we extend this analysis to
second order. We identify a minimal set of universal functions and give
their relation to matrix elements of higher dimension operators in the
effective theory. The zero recoil normalization conditions imposed on
some of these form factors are derived. Although in principle
straightforward, the analysis is tedious and involves considerable
technicalities of the heavy quark effective theory. The reader not
interested in these details is encouraged to proceed to
Sec.~\ref{sec:5}, where we summarize our results and illustrate them
for some specific cases of phenomenological relevance. In particular,
the corrections affecting the determination of $V_{cb}$ from exclusive
semileptonic $B$ decays are investigated. We also study the fictitious
limit of vanishing chromo-magnetic interaction, which leads to great
simplifications and might serve as an estimate of the dominant
corrections.

Based on the analysis for heavy mesons, the $1/m_Q^2$ corrections to
matrix elements between heavy baryons can readily be derived. We
discuss this subject in the following paper \cite{Baryons}.

\section{Parameters of the Effective Theory}
\label{sec:2}

The construction of the heavy quark effective theory (HQET) is based on
the observation that, in the limit $m_Q\gg\Lambda_{\rm QCD}$, the
velocity of a heavy quark is conserved with respect to soft processes.
It is then possible to remove the mass-dependent piece of the momentum
operator by a field redefinition. To this end, one introduces a field
$h_Q(v,x)$, which annihilates a heavy quark with velocity $v^\alpha$,
by \cite{Geor}
\begin{equation}\label{redef}
   h_Q(v,x) = e^{i m_Q v\cdot x}\,P_+(v)\,Q(x) \,,
\end{equation}
where $P_+(v)={1\over 2}(1+\rlap/v)$ is an on-shell projection
operator, and $Q(x)$ denotes the conventional quark field in QCD. If
$P^\alpha$ is the total momentum of the heavy quark, the new field
carries the residual momentum $k^\alpha=P^\alpha-m_Q v^\alpha$.

There is obviously some ambiguity associated with the construction of
HQET, since the heavy quark mass used in the definition of the field
$h_Q$ is not uniquely defined. In fact, for HQET to be consistent it is
only necessary that $k^\alpha$ be of order of $\Lambda_{\rm QCD}$, {\it
i.e.}, stay finite in the limit $m_Q\to\infty$. It is intuitively clear
that different choices for $m_Q$ must lead to the same answer for any
physical matrix element, and this can indeed be shown to be the case
\cite{AMM}. Yet it is advantageous to adopt a special choice for which
the resulting effective theory becomes particularly simple, in the
sense that there are no ``residual mass terms'' for the heavy quark and
the heavy quark expansion becomes a covariant derivative expansion.
This prescription provides a nonperturbative definition of the heavy
quark mass, which has been adopted implicitly in most previous analyses
based on HQET. It is important to realize, however, that so defined,
the mass $m_Q$ is a nontrivial parameter of the effective theory.

In the limit $m_Q\to\infty$, the effective Lagrangian for the strong
interactions of the heavy quark becomes \cite{Geor,Korn,Mann}
\begin{equation}\label{Leff}
   {\cal{L}}_{\rm HQET} = \bar h_Q\,i v\!\cdot\!D\,h_Q \,,
\end{equation}
where $D^\alpha=\partial^\alpha - i g_s t_a A_a^\alpha$ is the
gauge-covariant derivative. For finite $m_Q$, there appears in the
Lagrangian an infinite series of power corrections involving higher
dimension operators,
\begin{equation}\label{Lpower}
   {\cal{L}}_{\rm power} = {1\over 2 m_Q}\,{\cal{L}}_1
   + {1\over 4 m_Q^2}\,{\cal{L}}_2 + \cdots \,.
\end{equation}
Note that it is natural to expand in powers of $1/2 m_Q$ since, after
the field redefinition (\ref{redef}), $2 m_Q$ is the mass associated
with the heavy antiquark field which is integrated out \cite{Mann}.
Omitting an operator whose matrix elements vanish by the equation of
motion, the leading term in (\ref{Lpower}) is given by \cite{FGL}
\begin{equation}\label{L1}
   {\cal{L}}_1 = \bar h_Q\,(i D)^2 h_Q + Z(m_Q/\mu)\,
   \bar h_Q\,s_{\alpha\beta} G^{\alpha\beta} h_Q \,,
\end{equation}
where $s_{\alpha\beta}=-{i\over 2}\sigma_{\alpha\beta}$, and
$G^{\alpha\beta}=[i D^\alpha,i D^\beta]=i g_s t_a G_a^{\alpha\beta}$ is
the gluon field strength. In leading logarithmic approximation, the
renormalization factor for the chromo-magnetic moment operator is
\begin{equation}\label{Zfactor}
   Z(m_Q/\mu) = \Bigg[{\alpha_s(m_Q)\over\alpha_s(\mu)}\Bigg]^{9/\beta}
   \,;~ \beta = 33 - 2 n_f \,,
\end{equation}
where $n_f$ is the number of light quark flavors with mass below $m_Q$.
The kinetic term in (\ref{L1}) is not renormalized.

The purpose of the heavy quark expansion is to make the
$m_Q$-dependence of some hadronic quantity $A$ explicit by writing
\begin{equation}\label{expand}
   A(m_Q) = C_0(m_Q/\mu)\,A_0(\mu)
   + {1\over 2 m_Q}\,C_1(m_Q/\mu)\,A_1(\mu)
   + \cdots \,,
\end{equation}
in such a way that the coefficients $A_i(\mu)$ are universal,
$m_Q$-independent parameters, and $C_i(m_Q/\mu)$ are purely
perturbative coefficients, which dependent on $m_Q$ only via the
running of the strong coupling $\alpha_s(m_Q)$. The aim is to relate
$A_i$ to matrix elements of operators in HQET evaluated between the
eigenstates of the lowest order Lagrangian ${\cal{L}}_{\rm HQET}$. This
paper focuses on the ground state pseudoscalar and vector mesons, which
form a degenerate doublet under the heavy quark spin symmetry. These
mesons have the same velocity as the heavy quark which they contain.
Their common mass $M$, however, differs from the mass of the heavy
quarks by a finite amount $\bar\Lambda=M-m_Q$, which measures the
``mass'' carried by the light degrees of freedom. Because of the field
redefinition (\ref{redef}), it is this mass which governs the
$x$-dependence of states in the effective theory:
\begin{equation}\label{xdep}
   | M(x)\rangle_{\rm HQET} = e^{-i\bar\Lambda v\cdot x}
   | M(0)\rangle_{\rm HQET} \,.
\end{equation}
$\bar\Lambda$ is a universal parameter which can be defined in terms of
a matrix element of a higher dimension operator in HQET. Using the
equation of motion $i v\!\cdot\!D\,h_Q = 0$, which follows from the
effective Lagrangian ${\cal{L}}_{\rm HQET}$, it is easy to see that
\cite{AMM}
\begin{equation}
   \bar\Lambda = {\langle\,0\,|\,
    \bar q\,i v\!\cdot\!\overleftarrow{D}\,\Gamma\,h_Q\,| M(v)\rangle
    \over \langle\,0\,|\,\bar q\,\Gamma\,h_Q\,| M(v)\rangle} \,.
\end{equation}
Here $\Gamma$ is an appropriate Dirac matrix such that the currents
interpolate the heavy meson $M$. This relation shows that $\bar\Lambda$
is in fact a parameter describing the properties of the light degrees
of freedom in the background of the static color source provided by the
heavy quark. It turns out that this mass scale also enters the leading
power corrections to heavy meson form factors and determines the
canonical size of deviations from the infinite quark mass limit
\cite{Luke,GGW}. A recent analysis of $\bar\Lambda$ using QCD sum rules
predicts \cite{SR2}
\begin{equation}
   \bar\Lambda = 0.50\pm 0.07~{\rm GeV} \,.
\end{equation}

The eigenstates of ${\cal{L}}_{\rm HQET}$ differ from the states of
the full theory. In particular, their mass $M$ differs from the
physical masses of pseudoscalar or vector mesons by an amount of order
$1/m_Q$. These mass shifts are computable in HQET. The physical masses
$m_M$ obey a heavy quark expansion, which we write as $(m_M - m_Q) =
\bar\Lambda + \Delta m_M^2/2 m_Q + \cdots$. In the meson rest frame,
\begin{equation}\label{dm2}
   \Delta m_M^2 = {\langle M(v) |\,(- {\cal{L}}_1)\,| M(v)\rangle
              \over\langle M(v) |\,h_Q^\dagger h_Q\,| M(v)\rangle} \,.
\end{equation}
A convenient way to evaluate hadronic matrix elements in HQET
is by associating spin wave functions
\begin{equation}
   {\cal{M}}(v) = \sqrt{M}\,P_+(v) \cases{
    -\gamma_5 &; pseudoscalar meson $P$ \cr
    \rlap/\epsilon &; vector meson $V$ \cr}
\end{equation}
with the eigenstates of ${\cal{L}}_{\rm HQET}$ \cite{Falk,Bjor,AFF}.
These wave functions have the correct transformation properties under
boosts and heavy quark spin rotations. Here $\epsilon^\alpha$ denotes
the polarization vector of the vector meson. For reasons of simplicity
we shall often omit the argument $v$ in both $P_+$ and ${\cal{M}}$. We
note that ${\cal{M}}=P_+\,{\cal{M}}\,P_-$, where $P_\pm={1\over
2}(1\pm\rlap/v)$. Lorentz invariance allows one to write any matrix
element as a trace over these wave functions and appropriate Dirac
matrices. For the matrix elements in (\ref{dm2}) we define hadronic
parameters $\lambda_i$ by
\begin{eqnarray}\label{lamdef}
   \langle M |\,\bar h_Q\,(i D)^2 h_Q\,| M\rangle &=&
    - \lambda_1\,{\rm tr}\big\{\,\overline{\cal{M}}\,{\cal{M}}\,\big\}
    = 2 M\,\lambda_1 \,, \nonumber\\
   \langle M |\,\bar h_Q\,s_{\alpha\beta} G^{\alpha\beta}
    h_Q\,| M\rangle &=&
    - \lambda_2(\mu)\,{\rm tr}\big\{\,i\sigma_{\alpha\beta}\,
    \overline{\cal{M}}\,s^{\alpha\beta} {\cal{M}}\,\big\}
    = 2 d_M M\,\lambda_2(\mu) \,,
\end{eqnarray}
where $d_P=3$ for a pseudoscalar meson, and $d_V=-1$ for a vector
meson. The conservation of the vector current implies that, in the rest
frame, the matrix element in the denominator is given by $\langle M
|\,h_Q^\dagger h_Q\, | M\rangle = 2 M$. We thus have
\begin{equation}\label{Dm2}
   \Delta m_M^2 = - \lambda_1 - d_M\,Z(\mu)\,\lambda_2(\mu) \,.
\end{equation}
The universal parameters $\lambda_1$ and $\lambda_2$ are the analogs of
$\bar\Lambda$ at subleading order in the heavy quark expansion. They
are independent of $m_Q$. Whereas $\lambda_1$ is not renormalized,
$\lambda_2(\mu)$ depends on the renormalization scale in such a way
that the product $Z(\mu)\,\lambda_2(\mu)$ is scale-independent.

An estimate of the value of $\lambda_2$ can be obtained from the
measured mass splitting between the $B^*$ and $B$ mesons, assuming that
higher order corrections in the $B$ system are small. One finds
\begin{equation}
   m_{B^*}^2 - m_B^2 \approx \Delta m_{B^*}^2 - \Delta m_B^2
   = 4\lambda_2(m_b) \approx 0.48~{\rm GeV^2} \,,
\end{equation}
where the experimental value has been taken from Refs.~\cite{Bsplit}.
Using (\ref{Zfactor}) for the evolution of this parameter down to the
low energy scale $2\bar\Lambda\approx 1$ GeV, we obtain
\begin{equation}\label{lam2val}
   \lambda_2(2\bar\Lambda) \approx 0.15~{\rm GeV^2} \,.
\end{equation}
Unfortunately, it is not possible directly to relate the spin-symmetry
conserving parameter $\lambda_1$ to an observable quantity. Recently,
QCD sum rules have been used to compute both $\lambda_1$ and
$\lambda_2$ \cite{SR2}. The spin-symmetry breaking correction was found
in excellent agreement with experiment, $\lambda_2^{\rm s.r.} = 0.12\pm
0.02~{\rm GeV^2}$, and a rather large value for the spin-symmetry
conserving correction was obtained, $\lambda_1^{\rm s.r.}\approx
1~{\rm GeV^2}$. However, the sum rule analysis suggests that it might
be more appropriate to use an effective value of $\lambda_1$ in the $b$
and $c$ system which could be substantially smaller, even compatible
with zero. A measurement of $\lambda_1$ on a lattice could help to
clarify this issue.

\section{Meson Form Factors in the Effective Theory}
\label{sec:3}

Let us now review the analysis of current-induced transitions between
two heavy mesons to subleading order in HQET, as performed by Luke
\cite{Luke}. This will help to outline the general procedure and set up
the conventions we will need in Sec.~\ref{sec:4}. The aim is to
construct the heavy quark expansion (\ref{expand}) for matrix elements
of the type $\langle M'(v') |\,\bar Q'\,\Gamma\,Q\,| M(v)\rangle$,
where $\Gamma$ is an arbitrary Dirac matrix. In this case the universal
parameters are functions of the kinematic variable $w=v\cdot v'$, and
the perturbative coefficients, subsequently denoted by $C_j, c_j$ and
$c'_j$, depend on $w$ and both heavy quark masses. The current $\bar
Q'\,\Gamma\,Q$ has a short distance expansion in terms of operators of
the effective theory. It reads
\begin{eqnarray}\label{Jexp}
   \bar Q'\,\Gamma\,Q &\to& \sum_j C_j\,\bar h'\,\Gamma_j\,h
    \nonumber\\
   &&+ {1\over 2 m_Q} \sum_j c_j\,
    \bar h'\,\Gamma_j^\alpha\,i D_\alpha\,h
    + {1\over 2 m_{Q'}} \sum_j c'_j\,\bar h'\,
    (-i\overleftarrow{D}_{\!\alpha})\,\Gamma_j'^\alpha\,h + \cdots \,,
\end{eqnarray}
where we have abbreviated $h=h_Q(v)$ and $h'=h_{Q'}(v')$. The matrices
$\Gamma_j$ are in general different from $\Gamma$ and can depend on $v$
and $v'$. At tree level, however, one has
\begin{equation}
   \sum_j C_j\,\Gamma_j \to \Gamma \,,\qquad
   \sum_j c_j\,\Gamma_j^\alpha \to \Gamma\,\gamma^\alpha \,,\qquad
   \sum_j c'_j\,\Gamma_j'^\alpha \to \gamma^\alpha\Gamma \,.
\end{equation}

Using the trace formalism described in Sec.~\ref{sec:2}, matrix
elements of the leading term in (\ref{Jexp}) can be parameterized as
\begin{equation}\label{IWF}
   \langle M' |\,\bar h'\,\Gamma\,h\,| M \rangle = - \xi(w,\mu)\,
   {\rm tr}\big\{\, \overline{\cal{M}}'\,\Gamma\,{\cal{M}} \,\big\} \,,
\end{equation}
where we omit the velocity argument in the states and the wave
functions in order to simplify notation. It is to be understood that
quantities without a prime refer to the initial state meson $M$, while
primed quantities refer to the final state meson $M'$. Also, from now
on $m$ will designate a generic heavy quark mass. In general the form
factor $\xi$ could be some matrix-valued function of $v$ and $v'$, but
in this case the projection operators contained in the spin wave
functions restrict it to a scalar function of $w$. Eq.~(\ref{IWF})
implies that, to leading order in the heavy quark expansion, all matrix
elements of currents between pseudoscalar or vector mesons are
described by a single form factor, the Isgur-Wise function
\cite{Isgu,Falk,Bjor}. The kinematical information is contained in the
trace over spin wave functions. By evaluating the special case of
mesons with equal mass and velocity, one readily derives the zero
recoil normalization condition $\xi(1,\mu)=1$ as a consequence of the
conservation of the vector current.

At subleading order in (\ref{Jexp}) one encounters current operators
which contain a covariant derivative. Their matrix elements are
represented by the diagrams shown in Fig.~\ref{fig:1}(a) and can be
parameterized as
\begin{eqnarray}\label{ximudef}
   \langle M' |\,\bar h'\,\Gamma^\alpha\,i D_\alpha\,h\,| M\rangle
   &=& - {\rm tr}\big\{\, \xi_\alpha(v,v',\mu)\,\overline{\cal{M}}'
    \,\Gamma^\alpha {\cal{M}} \,\big\} \,, \nonumber\\
   \langle M' |\,\bar h'\,(-i\overleftarrow{D}_{\!\alpha})\,
    \Gamma'^\alpha h\,| M\rangle &=& - {\rm tr}\big\{\,
    \overline{\xi}_\alpha(v',v,\mu)\,\overline{\cal{M}}'\,
    \Gamma'^\alpha {\cal{M}} \,\big\} \,.
\end{eqnarray}
Note the interchange of the velocities in the second matrix element.
The most general decomposition of the universal form factor
$\xi_\alpha$ involves three scalar functions. Following
Ref.~\cite{Luke}, we define
\begin{equation}
   \xi_\alpha(v,v',\mu) = \xi_+(w,\mu)\,(v+v')_\alpha
   + \xi_-(w,\mu)\,(v-v')_\alpha - \xi_3(w,\mu)\,\gamma_\alpha \,.
\end{equation}
$T$-invariance of the strong interactions requires that these scalar
functions be real. Using (\ref{xdep}) and the fact that
$i\partial_\alpha(\bar h'\,\Gamma\,h) =
\bar h'\,i\overleftarrow{D}_{\!\alpha}\Gamma\,h +
\bar h'\,\Gamma\,i D_\alpha\,h$, one finds that
\begin{equation}
   \xi_-(w,\mu) = {\bar\Lambda\over 2}\,\xi(w,\mu) \,.
\end{equation}
This is where the parameter $\bar\Lambda$ enters the analysis.

The equation of motion, $i v\cdot D\,h=0$, yields an additional relation
among the scalar form factors. Taking into account that under the trace
$\xi_\alpha$ is sandwiched between projection operators, one obtains
\begin{equation}\label{xirela}
   P_-\,v^\alpha\xi_\alpha(v,v',\mu)\,P_-' = 0 \,.
\end{equation}
For the remainder of this paper we use the symbol ``$\hat =$'' for
relations such as this, which are true when sandwiched between the
projection operators provided by the meson wave functions. We thus
write $v^\alpha\xi_\alpha(v,v',\mu)\,\hat =\,0$. In terms of the scalar
functions this is equivalent to
\begin{equation}\label{xiirel}
   (w+1)\,\xi_+(w,\mu) - (w-1)\,\xi_-(w,\mu) + \xi_3(w,\mu) = 0 \,.
\end{equation}
We shall use this equation to eliminate $\xi_+$. In particular, it
follows that at zero recoil $2\,\xi_+(1,\mu) + \xi_3(1,\mu)=0$. This
relation has an interesting consequence, since it implies that
\begin{equation}
   \xi_\alpha(v,v,\mu)\,\hat =\,
   \big[ 2\,\xi_+(1,\mu) + \xi_3(1,\mu) \big]\,v_\alpha = 0 \,,
\end{equation}
showing that matrix elements of the higher dimension currents in
(\ref{Jexp}) vanish at zero recoil. This is the first part of Luke's
theorem \cite{Luke}. In its above form it is obvious that this result
is true to all orders in perturbation theory \cite{Ben}, since it does
not rely on the structure of the perturbative coefficients in
(\ref{Jexp}).

A second class of $1/m$ corrections comes from the presence of higher
dimension operators in the effective Lagrangian. Insertions of
operators of ${\cal{L}}_1$ in (\ref{Lpower}) into matrix elements of
the leading order currents represent corrections to the wave functions,
which appear since the eigenstates of ${\cal{L}}_{\rm HQET}$ are
different from the eigenstates of the full theory. The corresponding
diagrams are shown in Fig.~\ref{fig:1}(b). The relevant matrix elements
can be written as
\begin{eqnarray}\label{Adef}
   \langle M' |\,i\!\int\!{\rm d}x\,T\big\{\, J(0),{\cal{L}}_1(x)
    \,\big\} \,| M \rangle
    &=& - A_1(w,\mu)\,{\rm tr}\big\{\, \overline{\cal{M}}'\,
    \Gamma\,{\cal{M}} \,\big\} \nonumber\\
   &&- Z(m_Q/\mu)\,{\rm tr}\big\{\, A_{\alpha\beta}(v,v',\mu)\,
    \overline{\cal{M}}'\,\Gamma\,P_+\,s^{\alpha\beta}
    {\cal{M}} \,\big\} \,, \nonumber\\
   && \\
   \langle M' |\,i\!\int\!{\rm d}x\,T\big\{\, J(0),{\cal{L}}_1'(x)
    \,\big\} \,| M \rangle
    &=& - A_1(w,\mu)\,{\rm tr}\big\{\, \overline{\cal{M}}'\,
    \Gamma\,{\cal{M}} \,\big\} \nonumber\\
   &&- Z(m_{Q'}/\mu)\,{\rm tr}\big\{\,
    \overline{A}_{\alpha\beta}(v',v,\mu)\,\overline{\cal{M}}'\,
    s^{\alpha\beta} P_+'\,\Gamma\,{\cal{M}} \,\big\} \,. \nonumber
\end{eqnarray}
where $J=\bar h'\,\Gamma\,h$ is a lowest order current. Noting that
$v_\alpha P_+ s^{\alpha\beta}{\cal{M}}=0$, we write the
decomposition\footnote{Our functions are related to those defined in
Ref.~\cite{Luke} by $A_1=2\chi_1, A_2=-2\chi_2$ and $A_3=4\chi_3$.}
\begin{equation}\label{Adecomp}
   A_{\alpha\beta}(v,v',\mu) = A_2(w,\mu)\,
   (v'_\alpha\gamma_\beta - v'_\beta\gamma_\alpha)
   + A_3(w,\mu)\,i \sigma_{\alpha\beta} \,.
\end{equation}

The four independent functions $\xi_3$ and $A_i$, as well as the mass
parameter $\bar\Lambda$, suffice to describe the first order power
corrections to any matrix element of a heavy quark current between
ground state mesons. To get a picture of the structure of the
corrections let us for simplicity neglect radiative corrections. In
this case, there is a simple relation between the currents in HQET and
the current in the full theory. Consider now the power corrections
proportional to $1/m_Q$. They leave the wave function of the final
state meson unaffected, but change the simple structure of
${\cal{M}}(v)$. The part proportional to the on-shell projection
operator $P_+$ will be modified, and a component proportional to $P_-$
will be induced, representing the ``small component'' of the full wave
function. Hence
\begin{equation}\label{modwf}
   {\cal{M}}(v) \to P_+(v)\,L_+^M(v,v') + P_-(v)\,L_-^M(v,v') \,.
\end{equation}
The general form of $L_\pm^M$ is
\begin{eqnarray}\label{Ldecomp}
   L_+^P(v,v') &=& \sqrt{M}\,(-\gamma_5)\,L_1(w) \,, \nonumber\\
   L_+^V(v,v') &=& \sqrt{M}\,\Big[ \rlap/\epsilon\,L_2(w)
                + \epsilon\!\cdot\! v'\,L_3(w) \Big] \,, \nonumber\\
   L_-^P(v,v') &=& \sqrt{M}\,(-\gamma_5)\,L_4(w) \,, \nonumber\\
   L_-^V(v,v') &=& \sqrt{M}\,\Big[ \rlap/\epsilon\,L_5(w)
                + \epsilon\!\cdot\! v'\,L_6(w) \Big] \,.
\end{eqnarray}
The insertions of higher order terms from the effective Lagrangian in
(\ref{Adef}) obviously contribute to $L_+^M$ only. On the other hand,
in the absence of radiative corrections the matrix elements of the
higher dimension currents can be written in the form
\begin{equation}
   \langle M' |\,\bar h'\,\Gamma\,i\,\rlap/\!D\,h\,| M \rangle
   = - {\rm tr}\big\{\, \overline{\cal{M}}'\,\Gamma\,P_-
   \big[ \gamma_\alpha\,{\cal{M}}\,\xi^\alpha(v,v') \big] \,\big\} \,,
\end{equation}
where we have used (\ref{xirela}) to insert $P_-$ between $\Gamma$ and
$\gamma_\alpha$. Consequently, these corrections contribute to $L_-^M$
only. This is expected since $i\,\rlap/\!D\,h$ is proportional to the
small component of the full heavy quark spinor. By evaluating the
relevant traces one easily obtains
\begin{eqnarray}
   L_1 &=& A_1 + 2(w-1) A_2 + 3 A_3 \,, \nonumber\\
   L_2 &=& A_1 - A_3 \,, \nonumber\\
   L_3 &=& - 2 A_2 \,, \nonumber\\
   && \\
   L_4 &=& - \bar\Lambda\,\xi + 2\,\xi_3 \,, \nonumber\\
   L_5 &=& - \bar\Lambda\,\xi \,, \nonumber\\
   L_6 &=& - {2\over w+1}\,\big( \bar\Lambda\,\xi + \xi_3 \big) \,,
   \nonumber
\end{eqnarray}
and the complete matrix element becomes
\begin{eqnarray}\label{matrixel}
   \langle M' |\,\bar Q'\,\Gamma\,Q\,| M \rangle &=&
    - \xi(w)\,{\rm tr}\big\{\, \overline{\cal{M}}'\,\Gamma\,{\cal{M}}
    \,\big\} \nonumber\\
   &&- {1\over 2 m_Q}\,{\rm tr}\Big\{\, \overline{\cal{M}}'\,
    \Gamma\,\Big[ P_+\,L_+^M(v,v') + P_-\,L_-^M(v,v') \Big]
    \Big\} \\
   &&- {1\over 2 m_{Q'}}\,{\rm tr}\Big\{\, \Big[
    \overline{L}_+^{M'}(v',v)\,P_+' + \overline{L}_-^{M'}(v',v)\,P_-'
    \Big]\,\Gamma\,{\cal{M}} \,\Big\} + \cdots \,. \nonumber
\end{eqnarray}
This way of organizing the corrections reduces to a minimum the effort
required to compute the traces. If radiative corrections are taken into
account, it still suffices to define these six functions $L_i$, as long
as one stays with the leading logarithmic approximation for the
perturbative coefficients. The corresponding expressions are given in
Ref.~\cite{Sublea}, where the functions $L_i$ were called $\varrho_i$.

Let us evaluate (\ref{matrixel}) for the matrix elements of the vector
and axial vector currents, $V_\mu$ and $A_\mu$, between bottom and
charm mesons, which can be described completely in terms of fourteen
meson form factors $h_i(w)$. We define
\begin{eqnarray}\label{formfa}
   \langle D(v') |\,V_\mu\,| B(v) \rangle &=&
    \sqrt{m_B m_D}\,\Big[ h_+(w)\,(v+v')_\mu +
    h_-(w)\,(v-v')_\mu \Big] \,, \nonumber\\
   && \nonumber\\
   \langle D^*(v',\epsilon') |\,V_\mu\,| B(v) \rangle &=&
    i \sqrt{m_B m_{D^*}}\,\,h_V(w)\,
    \epsilon_{\mu\nu\alpha\beta}
    \,\epsilon'^{*\nu}\,v'^\alpha\,v^\beta \,, \nonumber\\
   && \nonumber\\
   \langle D^*(v',\epsilon') |\,A_\mu\,| B(v) \rangle &=&
    \sqrt{m_B m_{D^*}}\,\Big[ h_{A_1}(w)\,(w+1)\,
    \epsilon_\mu'^* \nonumber\\
   &&\qquad\qquad - h_{A_2}(w)\,\epsilon'^*\!\!\cdot\! v\,v_\mu
    - h_{A_3}(w)\,\epsilon'^*\!\!\cdot\! v\,v'_\mu \Big] \,, \nonumber\\
   && \\
   \langle D^*(v',\epsilon') |\,V_\mu\,| B^*(v,\epsilon) \rangle
    &=& \sqrt{m_{B^*} m_{D^*}}\,\bigg\{
    - \epsilon\!\cdot\!\epsilon'^*\,
    \Big[ h_1(w)\,(v+v')_\mu + h_2(w)\,(v-v')_\mu \Big] \nonumber\\
   &&\qquad\qquad\quad~ + h_3(w)\,\epsilon'^*\!\!\cdot\!v\,\epsilon_\mu
    + h_4(w)\,\epsilon\!\cdot\!v'\,\epsilon'^*_\mu \nonumber\\
   &&\qquad\qquad\quad~ - \epsilon\!\cdot\!v'\,
    \epsilon'^*\!\!\cdot\!v\,\Big[ h_5(w)\,v_\mu
    + h_6(w)\,v'_\mu \Big] \bigg\} \,, \nonumber\\
   && \nonumber\\
   \langle D^*(v',\epsilon') |\,A_\mu\,| B^*(v,\epsilon) \rangle
    &=& i \sqrt{m_{B^*} m_{D^*}}\,\epsilon_{\mu\nu\alpha\beta}\,
    \epsilon^\alpha\,\epsilon'^{*\beta}\,
    \Big[ h_7(w)\,(v+v')^\nu + h_8\,(v-v')^\nu \Big] \,. \nonumber
\end{eqnarray}
At leading order in the heavy quark expansion one finds that
\begin{equation}
   h_+ = h_V = h_{A_1} = h_{A_3} = h_1 = h_3 = h_4 = h_7 = \xi \,,
\end{equation}
while the remaining six form factors vanish. The expressions arising at
subleading order are given in Appendix~\ref{app:2}. Here we restrict
ourselves to three important cases, namely
\begin{eqnarray}\label{Luke2}
   h_+(w) &=& \xi(w) + \bigg({1\over 2 m_c} + {1\over 2 m_b}\bigg)\,
    L_1(w) \,, \nonumber\\
   && \nonumber\\
   h_1(w) &=& \xi(w) + \bigg({1\over 2 m_c} + {1\over 2 m_b}\bigg)\,
    L_2(w) \,, \\
   && \nonumber\\
   h_{A_1}(w) &=& \xi(w)
    + {1\over 2 m_b}\,\bigg[ L_1(w) - {w-1\over w+1}\,L_4(w) \bigg]
    + {1\over 2 m_c}\,\bigg[ L_2(w) - {w-1\over w+1}\,L_5(w) \bigg]
    \,. \nonumber
\end{eqnarray}
The conservation of the vector current in the limit $m_b=m_c$ implies
the zero recoil normalization conditions $h_+(1) = h_1(1) = 1$, from
which it follows that $L_1(1)=L_2(1)=0$, {\it i.e.\/} \cite{Luke}
\begin{equation}
   A_1(1,\mu) = A_3(1,\mu) = 0 \,.
\end{equation}
This is the second part of Luke's theorem, which is again true to all
orders in perturbation theory. It follows that
$A_{\alpha\beta}(v,v,\mu) \,\hat =\,0$, so that the matrix elements in
(\ref{Adef}) vanish at zero recoil.

In summary, Luke's theorem implies that the {\it matrix elements\/}
which describe the first order power corrections in HQET vanish at zero
recoil. It is important to realize that this does not imply that the
meson {\it form factors\/} are unaffected by $1/m$ corrections
\cite{Volker}. In fact, the theorem only applies for form factors which
are not kinematically suppressed as $v'\to v$. Besides $h_+$ and $h_1$
those are $h_{A_1}$ and $h_7$. The form factor $h_{A_1}$, which
according to (\ref{Luke2}) is indeed seen to be unaffected by first
order power corrections at zero recoil, plays an important role in the
determination of $V_{cb}$ from semileptonic decays \cite{MN}. It is one
of the purposes of the next section to investigate the second order
corrections to this form factor.

\section{Second Order Power Corrections}
\label{sec:4}

The analysis of higher order corrections in HQET makes use of the same
techniques as those developed above. The second order corrections can
be divided into three classes: corrections to the current, corrections
to the effective Lagrangian, and mixed corrections. We shall discuss
each of them separately below. In order to keep the presentation as
simple as possible we will often ignore radiative corrections; however,
we will always make clear how they could be incorporated into our
analysis.

\subsection{\bf Second Order Corrections to the Current}

At tree level, the expansion of the heavy quark current reads
[cf.~(\ref{Jexp})]
\begin{eqnarray}\label{Jexp2}
   \bar Q'\,\Gamma\,Q &\to& \bar h'\,\Gamma\,h
    + {1\over 2 m_Q}\,\bar h'\,\Gamma\,i\,\rlap/\!D\,h
    + {1\over 2 m_{Q'}}\,\bar h'\,(-i\overleftarrow{\,\rlap/\!D})\,
    \Gamma\,h \nonumber\\
   &&+ {1\over 4 m_Q^2}\,\bar h'\,\Gamma\,\gamma_\alpha v_\beta
    G^{\alpha\beta} h - {1\over 4 m_{Q'}^2}\,
    \bar h'\,\gamma_\alpha v'_\beta G^{\alpha\beta} \Gamma\,h
    \nonumber\\
   &&+ {1\over 4 m_Q m_{Q'}}\,\bar h'\,(-i\overleftarrow{\,\rlap/\!D})
      \,\Gamma\,i\,\rlap/\!D\,h + \cdots \,.
\end{eqnarray}
On dimensional grounds, the operators appearing at second order are
bilinear in the covariant derivative (recall that $G^{\alpha\beta} =
[i D^\alpha, i D^\beta]$). This remains true in the presence of
radiative corrections, although a number of additional operators are
induced. It thus suffices to consider the single hadronic matrix element
\begin{equation}\label{psidef}
   \langle M' |\,\bar h'\,(-i\overleftarrow{D}_{\!\alpha})\,
   \Gamma^{\alpha\beta}\,i D_\beta\,h\,| M\rangle =
   - {\rm tr}\big\{\, \psi_{\alpha\beta}(v,v',\mu)\,
   \overline{\cal{M}}'\,\Gamma^{\alpha\beta} {\cal{M}} \,\big\} \,,
\end{equation}
represented by the third diagram in Fig.~\ref{fig:2}(a). From now on we
shall always omit the $\mu$-dependence of the universal form factors
except in the equations which define them. Considering the complex
conjugate of the above matrix element, one finds that the form factor
must obey the symmetry relation
\begin{equation}\label{psisym}
   \overline{\psi}_{\beta\alpha}(v',v) = \psi_{\alpha\beta}(v,v') \,,
\end{equation}
which reduces the number of invariant functions to seven. It is
convenient to perform a decomposition into symmetric and antisymmetric
parts, $\psi_{\alpha\beta} = {1\over 2}[\psi^S_{\alpha\beta} +
\psi^A_{\alpha\beta}]$, and to define
\begin{eqnarray}
   \psi_{\alpha\beta}^S(v,v') &=& \psi_1^S(w)\,g_{\alpha\beta}
    + \psi_2^S(w)\,(v+v')_\alpha (v+v')_\beta
    + \psi_3^S(w)\,(v-v')_\alpha (v-v')_\beta \nonumber\\
   &&+ \psi_4^S(w)\,\Big[ (v+v')_\alpha\gamma_\beta
    + (v+v')_\beta\gamma_\alpha \Big] \,, \nonumber\\
   && \\
   \psi_{\alpha\beta}^A(v,v')
   &=& \psi_1^A(w)\,(v_\alpha v'_\beta - v'_\alpha v_\beta)
    + \psi_2^A(w)\,\Big[ (v-v')_\alpha\gamma_\beta
    - (v-v')_\beta\gamma_\alpha \Big] \nonumber\\
   &&+ i \psi_3^A(w)\,\sigma_{\alpha\beta} \,. \nonumber
\end{eqnarray}
As in (\ref{xirela}) one can use the equation of motion to derive
relations among the scalar form factors. It follows that under the trace
\begin{equation}
   v^\beta\psi_{\alpha\beta}(v,v')\,\hat =\,0 ,\qquad
   v'^\alpha\psi_{\alpha\beta}(v,v')\,\hat =\,0 \,.
\end{equation}
These conditions are equivalent because of (\ref{psisym}) and lead to
the three relations
\begin{eqnarray}\label{psirel}
   \psi_1^S + (w+1)\,\psi_2^S - (w-1)\,\psi_3^S - \psi_4^S +
    w\,\psi_1^A - \psi_2^A - \psi_3^A &=& 0 \,, \nonumber\\
   (w+1)\,\psi_2^S + (w-1)\,\psi_3^S - \psi_4^S -
    \psi_1^A + \psi_2^A &=& 0 \,, \nonumber\\
   (w+1)\,\psi_4^S + (w-1)\,\psi_2^A - \psi_3^A &=& 0 \,,
\end{eqnarray}
which reduce the number of independent functions to four.

One can use an integration by parts to relate (\ref{psidef}) to matrix
elements of operators containing two derivatives acting on the same
heavy quark field, which are represented by the first two diagrams in
Fig.~\ref{fig:2}(a). It follows that
\begin{eqnarray}\label{partint}
   \langle M' |\,\bar h'\,\Gamma^{\alpha\beta}\,
   i D_\alpha i D_\beta\,h\,| M\rangle &=&
    - {\rm tr}\big\{\, \psi_{\alpha\beta}(v,v')\,
    \overline{\cal{M}}'\,\Gamma^{\alpha\beta} {\cal{M}} \,\big\}
    \nonumber\\
   &&- \bar\Lambda\,(v-v')_\alpha\,{\rm tr}\big\{\,
    \xi_\beta(v,v')\,\overline{\cal{M}}'\,\Gamma^{\alpha\beta}
    {\cal{M}} \,\big\}
\end{eqnarray}
with $\xi_\beta$ as defined in (\ref{ximudef}). Matrix elements of
operators with both derivatives acting to the left can be obtained in a
similar way. In particular, we may derive from (\ref{partint}) the
matrix elements
\begin{eqnarray}\label{phis}
   \langle M' |\,\bar h'\,\Gamma\,(i D)^2 h\,| M\rangle &=&
    - \phi_0(w)\,{\rm tr}\big\{\,
    \overline{\cal{M}}'\,\Gamma\,{\cal{M}} \,\big\} \,, \nonumber\\
   \langle M' |\,\bar h'\,\Gamma^{\alpha\beta} G_{\alpha\beta}\,h\,
    | M\rangle &=& - {\rm tr}\big\{\, \phi_{\alpha\beta}(v,v')\,
    \overline{\cal{M}}'\,\Gamma^{\alpha\beta} {\cal{M}} \,\big\} \,,
\end{eqnarray}
the second of which is needed in (\ref{Jexp2}). Choosing the same
decomposition for $\phi_{\alpha\beta}$ as for $\psi_{\alpha\beta}^A$,
we find
\begin{eqnarray}
   \phi_0 &=& 2\psi_1^S + (w+1)\,\psi_2^S - (w-1)\,\psi_3^S -
    2\psi_4^S - \bar\Lambda^2 (w-1)\,\xi \,, \nonumber\\
   \phi_1 &=& \psi_1^A + \case{1}/{w+1} \big[
    \bar\Lambda^2 (w-1)\,\xi - 2\bar\Lambda\,\xi_3 \big] \,,
    \nonumber\\
   \phi_2 &=& \psi_2^A - \bar\Lambda\,\xi_3 \,, \nonumber\\
   \phi_3 &=& \psi_3^A \,.
\end{eqnarray}
We have already encountered matrix elements similar to (\ref{phis}) in
the discussion of mass shifts in Sec.~\ref{sec:2}, and from a
comparison with (\ref{lamdef}) we find the zero recoil conditions
\begin{eqnarray}\label{phinorm}
   \phi_0(1) &=& \lambda_1 \,, \nonumber\\
   \phi_3(1) &=& \lambda_2 \,, \nonumber\\
   \phi_1(1) - \phi_2(1) &=& -\case{1}/{3}\,\lambda_1
    + \case{1}/{2}\,\lambda_2 \,,
\end{eqnarray}
the last one being a consequence of the relations (\ref{psirel}), which
allow us also to express the form factors $\psi_i^S$ in terms of the
functions $\phi_i$. After some algebra we find
\begin{eqnarray}
   \psi_1^S &=& \phi_0 + w\,\phi_1 - {w\over w+1}\,(2\phi_2 + \phi_3)
    + \bigg({w-1\over w+1}\bigg)\,\bar\Lambda^2\,\xi \,, \nonumber\\
   && \nonumber\\
   \psi_2^S &=& -{1\over 2(w+1)}\,\big[ \phi_0 + (2w-1)\,\phi_1 \big]
    \nonumber\\
   &&+ {1\over 2(w+1)^2}\,\Big[ 2\phi_2 + (2w+3)\,\phi_3
    - (2-w)(w-1)\,\bar\Lambda^2\,\xi - 4(w-1)\,\bar\Lambda\,\xi_3
    \Big] \,, \nonumber\\
   && \\
   \psi_3^S &=& {1\over 2}\,(\hat\phi + \phi_1)
    - {1\over 4(w+1)}\,\Big[ 2\phi_2 + \phi_3 + 2w\,\bar\Lambda^2\,
    \xi \Big] \,, \nonumber\\
   && \nonumber\\
   \psi_4^S &=& {1\over w+1}\,\Big[ -(w-1)\,\phi_2 + \phi_3
    - (w-1)\,\bar\Lambda\,\xi_3 \Big] \,. \nonumber
\end{eqnarray}
We have introduced the function
\begin{equation}
   \hat\phi(w) = {1\over w-1}\,\Big[ \phi_0(w) + (w+2)\,\phi_1(w)
    - 3\,\phi_2(w) - \case{3}/{2}\,\phi_3(w) \Big] \,,
\end{equation}
which is nonsingular as $w\to 1$ because of (\ref{phinorm}).

The above relations allow us to prove a theorem which is the analog of
the first part of Luke's theorem:

\bigskip
\centerline{
\parbox{12.5truecm}
{{\it Theorem 1:}
At zero recoil, matrix elements of second order currents in the
heavy quark expansion can be expressed in terms of $\lambda_1$ and
$\lambda_2$.}
}
\bigskip

\noindent
For the proof we note that, to all orders in perturbation theory, the
relevant operators contain two covariant derivatives. Because of
(\ref{psidef}) and (\ref{partint}) the corresponding matrix elements at
zero recoil only involve $\psi_{\alpha\beta}(v,v)$ sandwiched between
projection operators. Using (\ref{phinorm}) we find that
\begin{equation}
   \psi_{\alpha\beta}(v,v)\,\hat =\,
   \big[ g_{\alpha\beta} - v_\alpha v_\beta\big]\,{\lambda_1\over 3}
   + i \sigma_{\alpha\beta}\,{\lambda_2\over 2} \,,
\end{equation}
which proves the theorem. Furthermore, we note that at tree level only
the last term in (\ref{Jexp2}) contributes at zero recoil, since
$v^\beta\phi_{\alpha\beta}(v,v)\,\hat =\,0$. The corresponding
corrections are of order $\lambda_i/m_Q m_{Q'}$.

\subsection{\bf Second Order Corrections to the Lagrangian}

Apart from operators whose matrix elements vanish by the equation of
motion, the most general form of the coefficient ${\cal{L}}_2$
appearing at second order in the expansion of the effective Lagrangian
in (\ref{Lpower}) contains two terms,
\begin{equation}
   {\cal{L}}_2 = Z_1(m_Q/\mu)\,\bar h\,v_\beta\,i D_\alpha
   G^{\alpha\beta} h + 2 Z_2(m_Q/\mu)\,\bar h\,s_{\alpha\beta}\,
   v_\gamma\,i D^\alpha G^{\beta\gamma} h \,.
\end{equation}
In leading logarithmic approximation the renormalization factors are
given by \cite{Lee}
\begin{eqnarray}
   Z_1(m_Q/\mu) &=& {55\over 9} - {46\over 9}\,
    \Bigg[{\alpha_s(m_Q)\over\alpha_s(\mu)}\Bigg]^{6/\beta} \,,
    \nonumber\\
   Z_2(m_Q/\mu) &=& {19\over 9} - {10\over 9}\,
   \Bigg[{\alpha_s(m_Q)\over\alpha_s(\mu)}\Bigg]^{9/\beta} \,.
\end{eqnarray}
These operators have the same Dirac structure as the operators in
${\cal{L}}_1$ in (\ref{L1}), and consequently their matrix elements are
of the same form as those of ${\cal{L}}_1$. In analogy to (\ref{Adef})
we thus define
\begin{eqnarray}\label{Bdef}
   \langle M' |\,i\!\int\!{\rm d}x\,T\big\{\, J(0),{\cal{L}}_2(x)
    \,\big\} \,| M \rangle
    &=& - Z_1\,B_1(w,\mu)\,{\rm tr}\big\{\, \overline{\cal{M}}'\,
    \Gamma\,{\cal{M}} \,\big\} \\
   &&- Z_2\,{\rm tr}\big\{\, B_{\alpha\beta}(v,v',\mu)\,
    \overline{\cal{M}}'\,\Gamma\,P_+\,s^{\alpha\beta}
    {\cal{M}} \,\big\} \,, \nonumber
\end{eqnarray}
and similarly for an insertion of ${\cal{L}}'_2$. As before,
$J=\bar h'\,\Gamma\,h$ denotes a lowest order current. The
corresponding diagrams are the first two shown in Fig.~\ref{fig:2}(b).
The decomposition of $B_{\alpha\beta}$ is of the same form as that for
$A_{\alpha\beta}$ in (\ref{Adecomp}). It involves two functions $B_2$
and $B_3$.

Another type of $1/m^2$ corrections comes from a double insertion of the
first order correction ${\cal{L}}_1$, as shown in the third and fourth
diagrams in Fig.~\ref{fig:2}(b). The corresponding matrix elements have
a more complicated structure. We define
\begin{eqnarray}\label{Cdef}
   \langle M' |\,\case{i^2}/{2}\!\int\!{\rm d}x{\rm d}y\,
   &T& \big\{\, J(0),{\cal{L}}_1(x),{\cal{L}}_1(y) \,\big\}
    \,| M \rangle \nonumber\\
   &=& - C_1(w,\mu)\,{\rm tr}\big\{\, \overline{\cal{M}}'\,
    \Gamma\,{\cal{M}} \,\big\}
    - Z\,{\rm tr}\big\{\, C_{\alpha\beta}(v,v',\mu)\,
    \overline{\cal{M}}'\,\Gamma\,P_+\,s^{\alpha\beta} {\cal{M}}
    \,\big\} \nonumber\\
   &&- Z^2\,{\rm tr}\big\{\, C_{\alpha\beta\gamma\delta}(v,v',\mu)\,
    \overline{\cal{M}}'\,\Gamma\,P_+\,s^{\alpha\beta} P_+\,
    s^{\gamma\delta} {\cal{M}} \,\big\} \,.
\end{eqnarray}
Again, the corresponding matrix elements with two insertions of
${\cal{L}}'_1$ can be obtained by conjugating the matrix elements as in
(\ref{Adef}). The decomposition of $C_{\alpha\beta}$ is the same as
that for $A_{\alpha\beta}$, involving two form factors $C_2$ and $C_3$.
The most general decomposition of the four-index object
$C_{\alpha\beta\gamma\delta}$ involves nine invariant functions, $C_4$
to $C_{12}$. They can be defined by
\begin{eqnarray}
   C_{\alpha\beta\gamma\delta}(v,v')
   &=& C_4(w)\,(g_{\alpha\gamma} g_{\beta\delta}
              - g_{\alpha\delta} g_{\beta\gamma})
     + C_5(w)\,\sigma_{\gamma\delta} \sigma_{\alpha\beta} \nonumber\\
   &+& C_6(w)\,(g_{\alpha\gamma} i\sigma_{\beta\delta}
              - g_{\beta\gamma} i\sigma_{\alpha\delta}
              - g_{\alpha\delta} i\sigma_{\beta\gamma}
              + g_{\beta\delta}i\sigma_{\alpha\gamma}) \nonumber\\
   &+& C_7(w)\,(v'_\gamma \gamma_\delta - v'_\delta \gamma_\gamma)
               (v'_\alpha \gamma_\beta - v'_\beta \gamma_\alpha)
               \nonumber\\
   &+& C_8(w)\,(g_{\alpha\gamma} v'_\beta v'_\delta
              - g_{\beta\gamma} v'_\alpha v'_\delta
              - g_{\alpha\delta} v'_\beta v'_\gamma
              + g_{\beta\delta} v'_\alpha v'_\gamma) \nonumber\\
   &+& C_9(w)\,(g_{\alpha\gamma} v'_\beta \gamma_\delta
              - g_{\beta\gamma} v'_\alpha \gamma_\delta
              - g_{\alpha\delta} v'_\beta \gamma_\gamma
              + g_{\beta\delta} v'_\alpha \gamma_\gamma) \nonumber\\
   &+& C_{10}(w)\,(g_{\alpha\gamma} \gamma_\beta v'_\delta
                 - g_{\beta\gamma} \gamma_\alpha v'_\delta
                 - g_{\alpha\delta} \gamma_\beta v'_\gamma
                 + g_{\beta\delta} \gamma_\alpha v'_\gamma) \nonumber\\
   &+& C_{11}(w)\,(i\sigma_{\alpha\gamma} v'_\beta \gamma_\delta
                 - i\sigma_{\beta\gamma} v'_\alpha \gamma_\delta
                 - i\sigma_{\alpha\delta} v'_\beta \gamma_\gamma
                 + i\sigma_{\beta\delta} v'_\alpha \gamma_\gamma)
                 \nonumber\\
   &+& C_{12}(w)\,(i\sigma_{\alpha\gamma} \gamma_\beta v'_\delta
                 - i\sigma_{\beta\gamma} \gamma_\alpha v'_\delta
                 - i\sigma_{\alpha\delta} \gamma_\beta v'_\gamma
                 + i\sigma_{\beta\delta} \gamma_\alpha v'_\gamma) \,.
\end{eqnarray}

Finally, there are corrections resulting from insertions of both
${\cal{L}}_1$ and ${\cal{L}}'_1$, as shown in the last diagram in
Fig.~\ref{fig:2}(b). They have the form
\begin{eqnarray}\label{Ddef}
   \langle M' |\,i^2\!\int\!{\rm d}x{\rm d}y\,
   &T& \big\{\, J(0),{\cal{L}}_1(x),{\cal{L}}'_1(y) \,\big\}
    \,| M \rangle \nonumber\\
   &=& - D_1(w,\mu)\,{\rm tr}\big\{\, \overline{\cal{M}}'\,
    \Gamma\,{\cal{M}} \,\big\} \nonumber\\
   &&- {Z\over 2}\,{\rm tr}\big\{\, D_{\alpha\beta}(v,v',\mu)\,
    \overline{\cal{M}}'\,\Gamma\,P_+\,s^{\alpha\beta} {\cal{M}}
    \,\big\} \nonumber\\
   &&- {Z'\over 2}\,{\rm tr}\big\{\,
    \overline{D}_{\alpha\beta}(v',v,\mu)\,\overline{\cal{M}}'\,
    s^{\alpha\beta} P_+'\,\Gamma\,{\cal{M}} \,\big\} \nonumber\\
   &&- Z Z'\,{\rm tr}\big\{\, D_{\alpha\beta\gamma\delta}(v,v',\mu)\,
    \overline{\cal{M}}'\,s^{\alpha\beta} P_+'\,\Gamma\,P_+\,
    s^{\gamma\delta} {\cal{M}} \,\big\} \,.
\end{eqnarray}
The form factor $D_{\alpha\beta}$ is again of the same form as
$A_{\alpha\beta}$ and involves two functions, $D_2$ and $D_3$. The
most general decomposition of the four-index object
$D_{\alpha\beta\gamma\delta}$ is similar to that of
$C_{\alpha\beta\gamma\delta}$. However, because of the symmetry of the
matrix element (\ref{Ddef}) this quantity has to obey the constraint
\begin{equation}
   D_{\alpha\beta\gamma\delta}(v,v') =
   \overline{D}_{\gamma\delta\alpha\beta}(v',v) ,
\end{equation}
which allows only seven independent functions, $D_4$ to $D_{10}$. We
choose the decomposition
\begin{eqnarray}
   D_{\alpha\beta\gamma\delta}(v,v')
   &=& D_4(w)\,(g_{\alpha\gamma} g_{\beta\delta}
              - g_{\alpha\delta} g_{\beta\gamma})
     + D_5(w)\,\sigma_{\gamma\delta} \sigma_{\alpha\beta} \nonumber\\
   &+& D_6(w)\,(g_{\alpha\gamma} i\sigma_{\beta\delta}
              - g_{\beta\gamma} i\sigma_{\alpha\delta}
              - g_{\alpha\delta} i\sigma_{\beta\gamma}
              + g_{\beta\delta}i\sigma_{\alpha\gamma}) \nonumber\\
   &+& D_7(w)\,(v'_\gamma \gamma_\delta - v'_\delta \gamma_\gamma)
               (v_\alpha \gamma_\beta - v_\beta \gamma_\alpha)
               \nonumber\\
   &+& D_8(w)\,(g_{\alpha\gamma} v_\beta v'_\delta
              - g_{\beta\gamma} v_\alpha v'_\delta
              - g_{\alpha\delta} v_\beta v'_\gamma
              + g_{\beta\delta} v_\alpha v'_\gamma) \nonumber\\
   &+& D_9(w)\,\Big[ g_{\alpha\gamma} v_\beta \gamma_\delta
              - g_{\beta\gamma} v_\alpha \gamma_\delta
              - g_{\alpha\delta} v_\beta \gamma_\gamma
              + g_{\beta\delta} v_\alpha \gamma_\gamma \nonumber\\
   && \phantom{D_9(w)}
              + g_{\alpha\gamma} \gamma_\beta v'_\delta
              - g_{\beta\gamma} \gamma_\alpha v'_\delta
              - g_{\alpha\delta} \gamma_\beta v'_\gamma
              + g_{\beta\delta} \gamma_\alpha v'_\gamma \Big]
              \nonumber\\
   &+& D_{10}(w)\,\Big[ v_\beta \gamma_\delta i\sigma_{\alpha\gamma}
              - v_\alpha \gamma_\delta i\sigma_{\beta\gamma}
              - v_\beta \gamma_\gamma i\sigma_{\alpha\delta}
              + v_\alpha \gamma_\gamma i\sigma_{\beta\delta}
              \nonumber\\
   && \phantom{D_{10}(w)}
              + i\sigma_{\alpha\gamma} \gamma_\beta v'_\delta
              - i\sigma_{\beta\gamma} \gamma_\alpha v'_\delta
              - i\sigma_{\alpha\delta} \gamma_\beta v'_\gamma
              + i\sigma_{\beta\delta} \gamma_\alpha v'_\gamma \Big] \,.
\end{eqnarray}
In total, twenty-five universal functions $B_i, C_i$ and $D_i$ are
necessary to parameterize the effects of second order corrections to
the effective Lagrangian of HQET. Unlike the corrections to the
current, there are no relations imposed on these form factors by the
equation of motion.

Before proceeding, we have to discuss an additional source of second
order corrections, which is related to the ones encountered above. As
discussed in Sec.~\ref{sec:2}, the mass $M$ in the wave functions that
we associate with the eigenstates of ${\cal{L}}_{\rm HQET}$ is
different from the physical mass $m_M$. It is the physical mass,
however, that appears in the normalization of matrix elements of the
vector current, which one uses to derive zero recoil conditions for
some of the universal form factors. At second order in the heavy quark
expansion one has to take into account this difference and perform a
mass renormalization of the wave function,
\begin{equation}
   {\cal{M}}(v) \to Z_M^{1/2}\,{\cal{M}}(v) ,\qquad
   Z_M^{1/2} = \sqrt{m_M\over M}
\end{equation}
in the first term in (\ref{matrixel}). This is compensated by a
counterterm
\begin{equation}
   - \Big[ 1 - Z_M^{1/2} Z_{M'}^{1/2} \Big]\,\xi(w)\,
   {\rm tr}\big\{\, \overline{\cal{M}}'\,\Gamma\,{\cal{M}} \,\big\}
   = \Bigg( {\Delta m_M^2\over 4 m_Q^2} +
   {\Delta m_{M'}^2\over 4 m_{Q'}^2} \Bigg)\,\xi(w)\,{\rm tr}\big\{\,
   \overline{\cal{M}}'\,\Gamma\,{\cal{M}} \,\big\} \,,
\end{equation}
which, according to (\ref{Dm2}), effectively adds $\lambda_1\,\xi$ to
$C_1$ and $\lambda_2\,\xi$ to $C_3$.

\subsection{\bf Combined Corrections to the Current and the Lagrangian}

The third and last type of $1/m^2$ corrections arises from the
combination of first order corrections both to the current and to the
Lagrangian, as shown in Fig.~\ref{fig:2}(c). The relevant structures are
\begin{eqnarray}\label{Edef}
   \langle M' |\,i\!\int\!{\rm d}x\, T\big\{\, \bar h'\,\Gamma^\gamma
   \,i D_\gamma\,h,{\cal{L}}_1(x) \,\big\} \,| M\rangle
   &=& - {\rm tr}\big\{\, E_\gamma(v,v',\mu)\,\overline{\cal{M}}'\,
    \Gamma^\gamma {\cal{M}} \,\big\} \nonumber\\
   &&- Z\,{\rm tr}\big\{\, E_{\gamma\alpha\beta}(v,v',\mu)\,
    \overline{\cal{M}}'\,\Gamma^\gamma P_+\,s^{\alpha\beta}
    {\cal{M}} \,\big\} \,, \nonumber\\
   && \\
   \langle M' |\,i\!\int\!{\rm d}x\, T\big\{\, \bar h'\,
   (-i\overleftarrow{D}_{\!\gamma})\,\Gamma\,h,{\cal{L}}_1(x)
   \,\big\} \,| M\rangle
   &=& - {\rm tr}\big\{\, E'_\gamma(v,v',\mu)\,\overline{\cal{M}}'\,
    \Gamma^\gamma {\cal{M}} \,\big\} \nonumber\\
   &&- Z\,{\rm tr}\big\{\, E'_{\gamma\alpha\beta}(v,v',\mu)\,
    \overline{\cal{M}}'\,\Gamma^\gamma P_+\,s^{\alpha\beta}
    {\cal{M}} \,\big\} \,. \nonumber
\end{eqnarray}
As previously, insertions of ${\cal{L}}'_1$ give rise to the conjugate
matrix elements. The form factor $E_\gamma$ may be parameterized as
\begin{equation}
   E_\gamma(v,v') = E_1(w)\,v_\gamma + E_2(w)\,v'_\gamma
   + E_3(w)\,\gamma_\gamma \,.
\end{equation}
The most general decomposition of $E_{\gamma\alpha\beta}$ involves
eight functions, which we define by
\begin{eqnarray}
   E_{\gamma\alpha\beta}(v,v')
   &=& (v'_\alpha\gamma_\beta - v'_\beta\gamma_\alpha)\,\Big[
    E_4(w)\,v_\gamma + E_5(w)\,v'_\gamma + E_6(w)\,\gamma_\gamma \Big]
    \nonumber\\
   &&+ i\sigma_{\alpha\beta}\,\Big[ E_7(w)\,v_\gamma
    + E_8(w)\,v'_\gamma + E_9(w)\,\gamma_\gamma \Big] \nonumber\\
   &&+ \Big\{ g_{\alpha\gamma}\,\Big[ E_{10}(w)\,v'_\beta
    + E_{11}(w)\,\gamma_\beta \Big] - (\alpha\leftrightarrow\beta)
    \Big\} \,.
\end{eqnarray}
The equation of motion implies $v^\gamma E_\gamma\,\hat=\,0$ and
$v^\gamma E_{\gamma\alpha\beta}\,\hat =\,0$, which is equivalent to
\begin{eqnarray}
   E_1 + w\,E_2 - E_3 &=& 0 \,, \nonumber\\
   E_4 + w\,E_5 + E_6 &=& 0 \,, \nonumber\\
   E_7 + w\,E_8 - E_9 &=& 0 \,.
\end{eqnarray}
Here we have used the fact that $v_\alpha P_+ s^{\alpha\beta}
{\cal{M}} = 0$.

We define form factors $E'_i(w)$ by identical decompositions.
In this case, the equation of motion leads to the relations
\begin{eqnarray}
   w\,E'_1 + E'_2 - E'_3 &=& 0 \,, \nonumber\\
   w\,E'_4 + E'_5 - E'_6 + E'_{11} &=& 0 \,, \nonumber\\
   w\,E'_7 + E'_8 - E'_9 &=& 0 \,.
\end{eqnarray}
These functions are not independent of $E_i$, however, the reason being
that the matrix elements in (\ref{Edef}) are related to each other by
an integration by parts. This relation has its subtleties, since
insertions of ${\cal{L}}_1$ renormalize the masses of the states in the
effective theory and, therefore, modify the $x$-dependence of the
``bare'' states in (\ref{xdep}). In addition, there is a contact term
arising from the action of the derivative on the $\theta$-functions in
the time-ordered product. We discuss these issues in
Appendix~\ref{app:3}. One finds that the differences $(E_i-E'_i)$ are
in fact computable in terms of form factors introduced earlier. The
relations are
\begin{eqnarray}\label{Wardid}
   E_\gamma - E'_\gamma &=& \bar\Lambda\,(v-v')_\gamma\,A_1
    + v_\gamma\,(\phi_0 - \lambda_1\,\xi) \,, \nonumber\\
   && \\
   E_{\gamma\alpha\beta} - E'_{\gamma\alpha\beta} &=&
    \bar\Lambda\,(v-v')_\gamma\,A_{\alpha\beta}
    - v_\gamma(v'_\alpha\gamma_\beta - v'_\beta\gamma_\alpha)\,\phi_2
    \nonumber\\
   &&+ i v_\gamma\sigma_{\alpha\beta}\,(\phi_3 - \lambda_2\,\xi) \,.
    \nonumber
\end{eqnarray}
In particular, it follows that $E'_i=E_i$ for $i=3,6,9,10,11$, and we
will choose these five functions as a basis. Then a convenient way of
writing the solution of the constraints imposed by the equation of
motion is
\begin{eqnarray}
   E_1 + w\,E_2 &=& w\,E'_1 + E'_2 = E_3 \,, \nonumber\\
   E_1 - E_2 &=& \bar\Lambda A_1 + w\,\widetilde{\phi}_0 \,,
    \nonumber\\
   E'_1 - E'_2 &=& - \bar\Lambda A_1 + \widetilde{\phi}_0 \,,
    \nonumber\\
   && \nonumber\\
   E_4 + w\,E_5 &=& - E_6 \,, \nonumber\\
   w\,E'_4 + E'_5 &=& E_6 - E_{11} \,, \nonumber\\
   E_4 + E_5 &=& - {1\over w+1}\, \Big[ E_{11} + w\,\phi_2
    - \bar\Lambda\,(w-1) A_2 \Big] \,, \nonumber\\
   E'_4 + E'_5 &=& - {1\over w+1}\, \Big[ E_{11} - \phi_2
    - \bar\Lambda\,(w-1) A_2 \Big] \,, \nonumber\\
   && \nonumber\\
   E_7 + w\,E_8 &=& w\,E'_7 + E'_8 = E_9 \,, \nonumber\\
   E_7 - E_8 &=& \bar\Lambda A_3 + w\,\widetilde{\phi}_3 \,,
    \nonumber\\
   E'_7 - E'_8 &=& - \bar\Lambda A_3 + \widetilde{\phi}_3 \,,
\end{eqnarray}
where we have introduced the nonsingular functions
\begin{equation}
   \widetilde{\phi}_0(w) = {\phi_0(w) - \lambda_1\,\xi(w)\over w-1} \,,
   \qquad
   \widetilde{\phi}_3(w) = {\phi_3(w) - \lambda_2\,\xi(w)\over w-1} \,.
\end{equation}
Consistency of the equations determining $E_{4,5}$ and $E'_{4,5}$
furthermore requires that, at zero recoil,
\begin{equation}\label{Erela}
   2 E_6(1) - E_{11}(1) = \phi_2(1) \,.
\end{equation}

The constraints imposed by the equation of motion allow us to prove a
second theorem:

\bigskip
\centerline{
\parbox{12.5truecm}
{{\it Theorem 2:}
Matrix elements describing the mixed first order corrections to the
current and to the Lagrangian vanish at zero recoil.}
}
\bigskip

\noindent
It follows from the fact that, under the traces,
\begin{eqnarray}
   E_\gamma(v,v)\, &\hat =& \,v_\gamma\,\Big[
    E_1(1) + E_2(1) - E_3(1) \Big] = 0 \,, \nonumber\\
   E_{\gamma\alpha\beta}(v,v)\, &\hat =& \,i\sigma_{\alpha\beta}\,
    v_\gamma\,\Big[ E_7(1) + E_8(1) - E_9(1) \Big] = 0 \,,
\end{eqnarray}
with identical relations for $E'_i$. Thus, at zero recoil only genuine
second order corrections to the current or to the Lagrangian contribute
to hadronic form factors that are not kinematically suppressed. The
conservation of the vector current in the limit of equal masses then
leads to relations between the universal functions which describe the
corrections to the Lagrangian, and the parameters $\lambda_1$ and
$\lambda_2$ which, according to Theorem~1, describe the corrections to
the current. These normalization conditions are the subject of the
following subsection.

\subsection{\bf Modified Wave Functions and Normalization\\
                Conditions at Zero Recoil}

In this section we have shown that at second order in the heavy quark
expansion a total of $4+25+5=34$ universal functions is necessary to
parameterize, respectively, the effects of corrections to the current,
of corrections to the effective Lagrangian, and of the combined
corrections to both. The richness of the structures that arise might
seem both impressive and frustrating, and the effort required to
compute the various traces is quite considerable. However, only certain
combinations of form factors appear in the final expression for any
hadronic matrix element, and it is time to organize our results in a
more transparent and convenient way by employing the concept of
modified wave functions introduced in Sec.~\ref{sec:3}. The corrections
proportional to $1/m_Q^2$ change the wave function for the initial
state meson, but leave the final state unaffected (and vice versa for
the terms proportional to $1/m_{Q'}^2$). Their effects can therefore be
accounted for as in (\ref{modwf}). On the other hand, the corrections
proportional to $1/m_Q m_{Q'}$ affect both mesons and can only be
accounted for by a combined wave function. We can thus extend
(\ref{matrixel}) to second order by writing
\begin{eqnarray}\label{master}
   \langle M' |\,\bar Q'\,\Gamma\,Q\,| M \rangle &=&
    - Z_M^{1/2} Z_{M'}^{1/2}\,\xi(w)\,{\rm tr}\big\{\,
    \overline{\cal{M}}'\,\Gamma\,{\cal{M}} \,\big\} \nonumber\\
   &&- {1\over 2 m_Q}\,{\rm tr}\Big\{\, \overline{\cal{M}}'\,
    \Gamma\,\Big[ P_+ \Big( L_+^M + {1\over 2 m_Q}\,\ell_+^M \Big)
    + P_- \Big( L_-^M + {1\over 2 m_Q}\,\ell_-^M \Big) \Big] \Big\}
    \nonumber\\
   &&- {1\over 2 m_{Q'}}\,{\rm tr}\Big\{\, \Big[ \Big(
    \overline{L}_+^{M'} + {1\over 2 m_{Q'}}\,\overline{\ell}_+^{M'}
    \Big) P'_+ + \Big( \overline{L}_-^{M'} + {1\over 2 m_{Q'}}\,
    \overline{\ell}_-^{M'} \Big) P'_- \Big]\,\Gamma\,{\cal{M}}
    \,\Big\} \nonumber\\
   &&- {1\over 4 m_Q m_{Q'}}\,{\rm tr}\Big\{\, \Gamma\,\Big[
    P_+\,m_{++}^{M M'} P'_+ + P_-\,m_{--}^{M M'} P'_- \\
   &&\phantom{ - {1\over 4 m_Q m_{Q'}}\,{\rm tr}\,\,\, }
    + P_+\,m_{+-}^{M M'} P'_- + P_-\,m_{-+}^{M M'} P'_+ \Big] \Big\}
    + \cdots \,, \nonumber
\end{eqnarray}
where we have performed the mass renormalization for the leading term.
Here a ``bar'' denotes Dirac conjugation combined with an exchange of
velocities, polarizations, and masses. The virtue of (\ref{master}) is
that it allows an interpretation in terms of large and small
components, reducing to a minimum the effort required to perform the
traces. The structure of $\ell^M$ is the same as that of $L_\pm^M$ in
(\ref{Ldecomp}) and involves six functions $\ell_i(w)$. The structure
of $m^{M M'}$ is more complicated and requires the introduction of
twenty-four functions $m_i(w)$. They are defined in
Appendix~\ref{app:1}.

Let us now discuss how the various second order corrections fit into
this pattern. We start with the corrections to the Lagrangian, which
according to (\ref{Bdef}) and (\ref{Cdef}) preserve the $P_+$
projectors for the initial and final state. Hence, the fifteen
universal functions $B_i$ and $C_i$ contribute to $\ell_+^M$ only and
appear in the three combinations $\ell_1, \ell_2$ and $\ell_3$.
Similarly, the functions $D_i$ contribute to $m_{++}^{M M'}$ and enter
in the combinations $m_1$ to $m_7$. For the discussion of the
corrections to the current we restrict ourselves to the operators in
(\ref{Jexp2}), which are obtained from tree level matching of QCD and
HQET. As explained in Sec.~\ref{sec:3}, one can identify
$i\,\rlap/\!D\,h$ with the small component of the full heavy quark
spinor, and those terms lead to $P_-$ projectors in the modified wave
functions. The last operator in (\ref{Jexp2}) contains two such terms
and consequently contributes to $m_{--}^{M M'}$ only. In fact, using
the equation of motion its matrix elements can be written as
\begin{equation}
   \langle M' |\,\bar h'\,(-i\overleftarrow{\,\rlap/\!D})\,\Gamma\,
   i\,\rlap/\!D\,h\,| M \rangle = - {\rm tr}\big\{\, \Gamma\,P_-
   \big[ \gamma^\beta {\cal{M}}\,\psi^{\alpha\beta}\,
   \overline{\cal{M}}'\,\gamma^\alpha \big]\,P'_- \,\big\} \,.
\end{equation}
By evaluating the bracket one readily computes the functions $m_8$ to
$m_{14}$, which appear in the parameterization of $m_{--}^{M M'}$.
Because $\bar h'\,\Gamma\,\gamma_\alpha v_\beta G^{\alpha\beta} h =
- \bar h'\,\Gamma\,i v\!\cdot\!D\,i\,\rlap/\!D\,h$, the other
second order currents in (\ref{Jexp2}) contain one small component and
thus contribute to $\ell_-^M$. To see this, we employ the equation of
motion to write
\begin{equation}
   \langle M' |\,\bar h'\,\Gamma\,\gamma_\alpha v_\beta G^{\alpha\beta}
   h\,| M \rangle = - {\rm tr}\big\{\, \overline{\cal{M}}'\,\Gamma\,
   P_- \big[ \gamma^\alpha {\cal{M}}\,v^\beta \phi^{\alpha\beta}
   \big] \,\big\} \,.
\end{equation}
The mixed corrections to the current and the Lagrangian have the same
structure, since
\begin{equation}
   \langle M' |\,i\!\int\!{\rm d}x\, T\big\{\, \bar h'\,\Gamma\,
   i\,\rlap/\!D\,h,{\cal{L}}_1(x) \,\big\} \,| M\rangle
   = - {\rm tr}\big\{\, \overline{\cal{M}}'\,\Gamma\,P_- \big[
   \gamma^\gamma {\cal{M}}\,E_\gamma + \gamma^\gamma P_+\,
   s^{\alpha\beta} {\cal{M}}\,E_{\gamma\alpha\beta} \big] \,\big\} \,.
\end{equation}
Thus, both $\phi_i$ and $E_i$ enter in the functions $\ell_4, \ell_5$
and $\ell_6$. Finally, the second matrix element in (\ref{Edef})
determines the functions $m_{15}$ to $m_{24}$, which appear in the
decompositions of $m_{+-}^{M M'}$ and $m_{-+}^{M M'}$.

The complete set of expressions for $\ell_i$ and $m_i$ is given in
Appendix~\ref{app:1}, and in Appendix~\ref{app:2} we compute the meson
form factors $h_i$ in terms of these functions. Let us now use these
results to derive the normalization conditions which follow from the
conservation of the vector current in the limit of equal heavy quark
masses, $m_{Q'} = m_Q$. It implies that at zero recoil
\begin{equation}\label{CVC}
   \langle M(v) |\,\bar Q\,\gamma_\mu\,Q\,| M(v) \rangle
   = 2 m_M\,v_\mu
\end{equation}
for both pseudoscalar and vector mesons, which in terms of the meson
form factors is equivalent to $h_+(1) = h_1(1) = 1$.
It is now important that we have performed a mass renormalization
in the first term in (\ref{master}), since $m_M$ in (\ref{CVC}) is the
physical meson mass. Using the normalization of the Isgur-Wise function
and Luke's theorem, we find from Appendix~\ref{app:2} in the equal mass
limit
\begin{eqnarray}
   h_+(1) &=& 1 + {1\over 4 m_Q^2}\,\Big[ 2\ell_1(1) + m_1(1)
    - m_8(1) \Big] + \cdots \,, \nonumber\\
   h_1(1) &=& 1 + {1\over 4 m_Q^2}\,\Big[ 2\ell_2(1) + m_4(1) + m_5(1)
    - m_{11}(1) - m_{12}(1) \Big] + \cdots \,. \nonumber
\end{eqnarray}
Setting the coefficients of the second order terms to zero we obtain
two conditions, which at tree level may be written as
\begin{eqnarray}\label{zerorec}
   2 B_1(1) + 2 C_1(1) + D_1(1) - 3 \Big[ 2 C_4(1) + D_4(1)
    + 2 C_5(1) + D_5(1) \Big] &=& - \lambda_1 \,, \nonumber\\
   && \\
   2 B_3(1) + 2 C_3(1) + D_3(1) - 2 \Big[ 2 C_5(1) + D_5(1)
    + 2 C_6(1) + D_6(1) \Big] &=& - \lambda_2 \,. \nonumber
\end{eqnarray}
More restrictive relations could be derived by including renormalization
effects and requiring that the logarithmic dependence on $m_Q$ be the
same on both sides of (\ref{zerorec}). The results of such an analysis
will be presented elsewhere.

\section{Applications and Summary}
\label{sec:5}

Let us summarize the main results of our analysis. In total,
thirty-four universal functions appear in second order of the heavy
quark expansion of meson form factors. We have proved two theorems
stating that, at zero recoil, the leading meson form factors do not
receive contributions from mixed corrections to the current and the
Lagrangian, and that the corrections to the current can be expressed in
terms of $\lambda_1$ and $\lambda_2$. The number of universal functions
is strongly reduced if one ignores radiative corrections and only
considers the phenomenologically interesting cases of $P\to P$ and
$P\to V$ transitions induced by a vector or axial current. Then all
matrix elements can be parameterized in terms of $\ell_1$ to $\ell_6$
and the five combinations $(m_1-m_8)$, $(m_2+m_9)$, $(m_3-m_{10})$,
$(m_{16}+m_{18})$ and $(m_{17}-m_{19})$. This can be seen from the
relations given in Appendix~\ref{app:2}.

In the following paragraphs we apply our results to semileptonic $B$
decays and give estimates for some of the second order corrections. We
also discuss the corrections to Luke's theorem, which arise at second
order. For simplicity, we shall ignore radiative corrections.

\subsection{\bf Elastic Form Factors and $B\to D\,\ell\,\nu$ Decays}

As pointed out in the introduction, the universal form factors of HQET
describe the properties of the light degrees of freedom in the
background of the color field of the heavy quark. From this point of
view, the Isgur-Wise function is the elastic form factor that describes
the overlap of the wave functions of the light degrees of freedom in
the initial and final mesons moving at velocities $v$ and $v'$. The
normalization of $\xi(w)$ at zero recoil reflects the complete overlap
of the configurations of the light constituents in two infinitely heavy
mesons with the same velocity. If finite-mass corrections are taken
into account, the overlap decreases. In HQET the corresponding
corrections are described by the functions $L_i$ and $\ell_i$, which
represent the corrections to the wave function of a pseudoscalar
($i=1$) or a vector meson ($i=2$). At zero recoil, the first order
corrections vanish, and using the expression for $h_+$ and $h_1$ from
Appendix~\ref{app:2} we obtain at second order
\begin{eqnarray}\label{elast}
   \langle D(v) |\,V_\mu\,| B(v) \rangle &=& 2 \sqrt{m_B m_D}\,v_\mu\,
    \Big\{ 1 + (\varepsilon_c - \varepsilon_b)^2 \ell_1(1) + \cdots
    \Big\} \,, \nonumber\\
   && \\
   \langle D^*(v) |\,V_\mu\,| B^*(v) \rangle &=&
    - 2 \sqrt{m_{B^*} m_{D^*}}\,\epsilon\!\cdot\epsilon'^*\,v_\mu\,
    \Big\{ 1 + (\varepsilon_c - \varepsilon_b)^2 \ell_2(1) + \cdots
    \Big\} \,, \nonumber
\end{eqnarray}
where $\varepsilon_Q=1/2 m_Q$.

In the nonrelativistic constituent quark model, the $m_Q$-dependence of
the overlap integral comes from the $m_Q$-dependence of the reduced
mass of the light constituent quark, $m_q^{\rm red} =
m_q m_Q / (m_Q + m_q)$. For an estimate of $\ell_i(1)$ we use the
wave functions of the ISGW model \cite{ISGW} to obtain
\begin{equation}\label{redmass}
   \ell_1(1) = \ell_2(1) = -3 m_q^2 \approx -0.75~{\rm GeV^2} \,.
\end{equation}
For the numerical estimate we have identified the constituent mass of
the light quark with $\bar\Lambda$, since $m_M=m_Q+m_q$ in the ISGW
model.

The matrix element of the vector current between a $B$ and a $D$ meson
enters the theoretical description of the decay rate for the
semileptonic process $B\to D\,\ell\,\nu$. After contraction with the
leptonic current a combination of the form factors $h_+$ and $h_-$
appears \cite{Volker},
\begin{eqnarray}
   {{\rm d}\Gamma(B\to D\,\ell\,\nu)\over{\rm d}w} &=&
    {G_F^2\,|\,V_{cb}|^2\over 48\pi^3}\,m_D^3\,(m_B+m_D)^2\,
    (w^2-1)^{3/2} \nonumber\\
   &&\times \Big | h_+(w) - \sqrt{S}\,h_-(w) \Big |^2 \,,
\end{eqnarray}
where $S=\big({m_B-m_D\over m_B+m_D}\big)^2\approx 0.23$ is the
Voloshin-Shifman factor \cite{Volo}. At leading order in the heavy
quark expansion the form factor is normalized at zero recoil, offering
the possibility of a reliable determination of $V_{cb}$ for $w\agt 1$,
provided that the corrections to the infinite quark mass limit are
small. The first order power corrections are indeed suppressed by the
Voloshin-Shifman factor and have been estimated to be $\approx +2\%$
\cite{Sublea}. Including the second order corrections, we find from
Appendix~\ref{app:2}
\begin{eqnarray}
   h_+(1) - \sqrt{S}\,h_-(1) &=&
    1 - (\varepsilon_c + \varepsilon_b)\,S\,L_4(1)
    + (\varepsilon_c - \varepsilon_b)^2
    \big[ \ell_1(1) - \ell_4(1) \big]
    + 4 \varepsilon_c \varepsilon_b\,S\,\bar\Lambda \nonumber\\
   &\approx& 1 - 0.7\% - 1.3\% \times
    \bigg[{\ell_4(1)\over\bar\Lambda^2}\bigg] \,,
\end{eqnarray}
where we have used the heavy quark masses $m_c=1.5$ GeV and $m_b=4.8$
GeV, the constituent quark model estimate (\ref{redmass}), and the QCD
sum rule results $\bar\Lambda\approx 0.5$ GeV and $L_4(1)\approx
-\bar\Lambda/3$ \cite{Sublea}. For simplicity, the radiative
corrections to $h_+$ and $h_-$ have been neglected. We conclude that,
unless the coefficient $\ell_4(1)$ were unusually large, both the first
and second order power corrections are small. Although not protected by
Luke's theorem, the decay $B\to D\,\ell\,\nu$ thus allows for a
reliable measurement of $V_{cb}$.

\vfil\eject

\subsection{\bf Determination of $V_{cb}$ from $B\to D^*\ell\,\nu$
Decays}

It has been observed in Refs.~\cite{Volo,MN} that semileptonic $B$
decays into $D^*$ vector mesons offer an almost model-independent
measurement of $V_{cb}$, since the $1/m_Q$ corrections to the decay
rate vanish at zero recoil. In terms of the meson form factors one
finds
\begin{equation}
   \lim_{w\to 1} {1\over\sqrt{w^2-1}}\,
   {{\rm d}\Gamma(B\to D^*\ell\,\nu)\over{\rm d}w}
   = {G_F^2\,|\,V_{cb}|^2\over 4\pi^3}\,m_{D^*}^3\,(m_B-m_{D^*})^2\,
   \big | h_{A_1}(1) \big |^2 \,,
\end{equation}
and $h_{A_1}(1)$ is protected by Luke's theorem \cite{Luke}. Thus the
determination of $V_{cb}$ from an extrapolation of the spectrum to
$w\agt 1$ is model-independent up to terms of order $1/m^2$. From
Appendix~\ref{app:2} we obtain at second order
\begin{equation}\label{hA1}
   h_{A_1}(1) = 1 + (\varepsilon_c - \varepsilon_b)\,
   \big[ \varepsilon_c\,\ell_2(1) - \varepsilon_b\,\ell_1(1) \big]
   + \varepsilon_c \varepsilon_b\,\Delta \,,
\end{equation}
where
\begin{eqnarray}
   \Delta &=& \ell_1(1) + \ell_2(1) + m_2(1) + m_9(1) \nonumber\\
   &=& \case{4}/{3}\,\lambda_1 + 2\,\lambda_2 + 4\,\Big[
    D_4(1) + 2 D_5(1) + D_6(1) \Big] \,.
\end{eqnarray}
Using (\ref{redmass}), the first correction in (\ref{hA1}) is estimated
to be $-3\bar\Lambda^2(\varepsilon_c-\varepsilon_b)^2 \approx -3.9\%$.
Concerning the second term we observe that, except for $\lambda_1$ and
$\lambda_2$, the coefficient $\Delta$ depends only on form factors
which arise from a double insertion of the chromo-magnetic moment
operator of ${\cal{L}}_1$ in (\ref{L1}). We shall argue below that
these terms are expected to be very small. Neglecting them, and using
(\ref{lam2val}) as well as the sum rule estimate $\lambda_1\approx 1$
GeV$^2$ \cite{SR2}, we obtain $\varepsilon_c
\varepsilon_b\,\Delta\approx 5.7\%$, and thus
\begin{equation}\label{hA1val}
   h_{A_1}(1) - 1 \approx 2\% \,.
\end{equation}
The main uncertainty in this estimate arises from the uncertainty in
$\lambda_1$, as discussed at the end of Sec.~\ref{sec:2}. In the
extreme case $\lambda_1=0$ we would obtain $h_{A_1}(1) - 1 \approx
-3\%$ instead of (\ref{hA1val}). However, in any case the second
order correction is small because of a partial cancellation of the two
terms in (\ref{hA1}), suggesting that the theoretical uncertainty in
this method of extracting $V_{cb}$ is less than a few percent.

It has been claimed in Ref.~\cite{Ball} that QCD sum rules would
predict a second order correction to $h_{A_1}(1)$ of as much as
$-10\%$.\footnote{It has been pointed out in Ref.~\cite{Sublea} that
the argument given in Ref.~\cite{Ball} has no theoretical foundation.}
In view of our estimate (\ref{hA1val}) this assertion seems
unacceptable. Even for $\lambda_1=0$ it would imply that $\ell_1(1)$
and $\ell_2(1)$ would have to exceed the quark model prediction
(\ref{redmass}) by a factor of three.

\subsection{\bf Second Order Corrections to Luke's Theorem}

At the end of Sec.~\ref{sec:3}, we discussed the fact that Luke's theorem
protects the meson form factors $h_+, h_{A_1}, h_1$, and $h_7$ from
first order power corrections at zero recoil. Although our results show
that there is no such nonrenormalization theorem at second order, the
structure of the $1/m^2$ corrections to these four form factors is
particularly simple and allows for a semi-quantitative estimate. At
zero recoil, the expression for $h_7$ is
\begin{equation}
   h_7(1) = 1 + (\varepsilon_c - \varepsilon_b)^2 \ell_2(1)
   + \varepsilon_c \varepsilon_b\,\Delta'
\end{equation}
with
\begin{equation}
   \Delta' = \case{4}/{3}\,\lambda_1 - 2\,\lambda_2 + 4\,
   \big[ D_4(1) - D_6(1) \big] \,.
\end{equation}
The other three form factors have been given in (\ref{elast}) and
(\ref{hA1}). We observe that there is always a correction involving
$\ell_1(1)$ or $\ell_2(1)$, depending on whether one deals with a
pseudoscalar or a vector meson, respectively. Using the quark model
estimate (\ref{redmass}), this term becomes approximately $-4\%$. It
smallness naturally results from the squared difference $(\varepsilon_c
- \varepsilon_b)^2$. In addition, for $h_{A_1}$ and $h_7$ there is a
term proportional to $\varepsilon_c \varepsilon_b$, which depends on
the mass parameters $\lambda_1$ and $\lambda_2$ as well as on form
factors arising from a double insertion of the chromo-magnetic moment
operator. Neglecting these latter terms, this correction can be
estimated based on a model calculation of $\lambda_1$, since
$\lambda_2$ is known from the experimentally observed mass splitting
between vector and pseudoscalar mesons. QCD sum rules predict that
$\lambda_1$ is positive, and the corresponding correction tends
to cancel the terms proportional to $\ell_i$, which are negative. As a
result, the form factors $h_{A_1}$ and $h_7$ can only receive small
$1/m^2$ corrections at zero recoil.

\subsection{\bf Limit of Vanishing Chromo-Magnetic Interaction}

Detailed QCD sum rule analyses of the universal functions that appear
at order $1/m$ in the heavy quark expansion show that the form factors
$A_2$ and $A_3$, which arise from the insertion of the chromo-magnetic
moment operator in ${\cal{L}}_1$, are much smaller than the other two
functions, $A_1$ and $\xi_3$ \cite{Sublea,Yossi}. The coarse pattern of
the $1/m$ corrections can be well described by setting $A_2$ and $A_3$
to zero, corresponding to the fictitious limit of vanishing field
strength, $G^{\alpha\beta}\to 0$. Let us see what kind of
simplifications the same approximation implies at order $1/m^2$.

We start with the corrections to the current. In the limit
$G^{\alpha\beta}\to 0$ the functions $\phi_1$, $\phi_2$ and $\phi_3$
vanish, and according to (\ref{phinorm}) this implies the vanishing of
$\lambda_1$ and $\lambda_2$. It then follows that $\phi_0 =
(w-1)\,\hat\phi$ vanishes at zero recoil, and all corrections to the
current can be described by the single function $\hat\phi$. Similar
simplifications occur for the corrections to the Lagrangian. Here all
universal functions except $B_1, C_1$, and $D_1$ vanish in the limit
$G^{\alpha\beta}\to 0$. At tree level, one obtains from (\ref{zerorec})
the zero recoil condition
\begin{equation}
   B_1(1) + C_1(1) + \case{1}/{2}\,D_1(1) = 0 \,. \qquad
   (G^{\alpha\beta}\to 0)
\end{equation}
Finally, the combined corrections to the current and to the Lagrangian
are entirely parameterized by the form factor $E_3$, since $E_6$, $E_9$,
$E_{10}$, and $E_{11}$ vanish in the limit of vanishing field strength.

In the fictitious limit of vanishing chromo-magnetic interaction, the
set of thirty-four universal form factors is thus reduced to only five
functions, a combination of which vanishes at zero recoil. Although we
are aware of the fact that such an approximation can only give us a
very simplified picture, we still believe that it might be useful for
an analysis of the structure of the dominant terms. The expressions
arising for the functions $\ell_i$ and $m_i$ in this limit can readily
be obtained from the general formulas given in Appendix~\ref{app:1}.

\subsection{\bf Summary}

Using the heavy quark effective theory, we have performed the expansion
of matrix elements of heavy quark currents between pseudoscalar or
vector mesons up to second order in inverse powers of the heavy quark
masses. The general description of the power corrections arising at
order $1/m^2$ involves a set of thirty-four Isgur-Wise form factors,
which are universal, $m_Q$-independent functions of the kinematic
variable $w=v\cdot v'$. These form factors are defined in terms of
matrix elements of higher dimension operators in the effective theory.

Apart from some normalization conditions imposed by vector current
conservation, the universal functions are hadronic quantities which
cannot yet be predicted from first principles. Nevertheless, we have
argued that in certain cases of phenomenological interest the $1/m^2$
corrections are parameterically suppressed. In particular, the
corrections to the semileptonic decay rates for $B\to D\,\ell\,\nu$ and
$B\to D^*\ell\,\nu$ at zero recoil are estimated to be small, not
exceeding a few percent. Our results thus support the usefulness of the
heavy quark symmetries for an accurate determination of the weak mixing
parameter $V_{cb}$ from these decay modes.

Although the structure of second order corrections to various decay
rates is quite complex, we believe that a classification in terms of
universal form factors is still a useful concept. In particular, this
might provide a framework in which to analyze various models. For
instance, we have shown that the second order corrections to elastic
form factors arising from the $m_Q$-dependence of the reduced mass of
the light constituent quark in a nonrelativistic quark model are
accounted for by our functions $\ell_1$ and $\ell_2$, and an estimate
of the effect gives $\ell_1(1)\approx\ell_2(1) \approx-0.75\,{\rm
GeV^2}$. This information can then be used to predict corrections to
other form factors, whose dependence on $\ell_1$ and $\ell_2$ is known
from heavy quark symmetry. We have also suggested that, for an estimate
of the dominant corrections, one might consider the limit of vanishing
chromo-magnetic interaction, in which only five of the thirty-four
universal form factors remain. The usefulness of such an approximation
is supported by QCD sum rule calculations of the form factors appearing
at order $1/m$ in the heavy quark expansion.

The analysis presented here for mesons can straightforwardly be
extended to other hadrons containing a single heavy quark. The
particularly interesting case of the $\Lambda$ baryons is discussed in
Ref.~\cite{Baryons}.

\acknowledgements
It is a pleasure to thank Nathan Isgur, Peter Lepage, Michael Peskin,
and Stephen Sharpe for very helpful discussions. M.N. gratefully
acknowledges financial support from the BASF Aktiengesellschaft and
from the German National Scholarship Foundation. This work was also
supported by the Department of Energy, contract DE-AC03-76SF00515.

\newpage

\appendix{Computation of the Modified Wave Functions}
\label{app:1}

According to (\ref{Ldecomp}), the general structure of the modified
wave functions $\ell_\pm^M$ introduced in (\ref{master}) is
\begin{eqnarray}
   \ell_+^P &=& \sqrt{M}\,(-\gamma_5)\,\ell_1 \,, \nonumber\\
   \ell_+^V &=& \sqrt{M}\,\Big[ \rlap/\epsilon\,\ell_2
    + \epsilon\!\cdot\! v'\,\ell_3 \Big] \,, \nonumber\\
   \ell_-^P &=& \sqrt{M}\,(-\gamma_5)\,\ell_4 \,, \nonumber\\
   \ell_-^V &=& \sqrt{M}\,\Big[ \rlap/\epsilon\,\ell_5
    + \epsilon\!\cdot\! v'\,\ell_6 \Big] \,.
\end{eqnarray}
The coefficients $\ell_i$ are functions of $w=v\cdot v'$. Similarly, we
choose the following decomposition for the product wave functions
$m_{++}^{M M'}$:
\begin{eqnarray}
   m_{++}^{P P} &=& \sqrt{M M'}\,\,m_1\,(-\gamma_5)\,\gamma_5
    = - \sqrt{M M'}\,\,m_1 \,, \nonumber\\
   m_{++}^{P V} &=& \sqrt{M M'}\,
    \Big[ m_2\,(-\gamma_5)\,\rlap/\epsilon'^*
    + m_3\,(-\gamma_5)\,\epsilon'^*\!\!\cdot\!v \Big] \,, \nonumber\\
   m_{++}^{V P} &=& \sqrt{M M'}\,
    \Big[ m'_2\,\rlap/\epsilon\,\gamma_5
    + m'_3\,\epsilon\!\cdot\!v'\,\gamma_5 \Big]
    = \overline{m}_{++}^{P V} \,, \nonumber\\
   m_{++}^{V V} &=& \sqrt{M M'}\, \Big[
    m_4\,\rlap/\epsilon\,\rlap/\epsilon'^*
    + m_5\,\epsilon\!\cdot\!\epsilon'^*
    + m_6\,\rlap/\epsilon\,\epsilon'^*\!\!\cdot\!v
    + m'_6\,\rlap/\epsilon'^*\,\epsilon\!\cdot\!v'
    + m_7\,\epsilon\!\cdot\!v'\,\epsilon'^*\!\!\cdot\!v \Big] \,.
\end{eqnarray}
An identical decomposition with functions $m_8$ to $m_{14}$ applies for
$m_{--}^{M M'}$. Finally, we define
\begin{eqnarray}
   m_{+-}^{P P} &=& - \sqrt{M M'}\,\,m_{15} \,, \nonumber\\
   m_{+-}^{P V} &=& \sqrt{M M'}\,
    \Big[ m_{16}\,(-\gamma_5)\,\rlap/\epsilon'^*
    + m_{17}\,(-\gamma_5)\,\epsilon'^*\!\!\cdot\!v \Big] \,, \\
   m_{+-}^{V P} &=& \sqrt{M M'}\,
    \Big[ m_{18}\,\rlap/\epsilon\,\gamma_5
    + m_{19}\,\epsilon\!\cdot\!v'\,\gamma_5 \Big] \,, \nonumber\\
   m_{+-}^{V V} &=& \sqrt{M M'}\, \Big[
    m_{20}\,\rlap/\epsilon\,\rlap/\epsilon'^*
    + m_{21}\,\epsilon\!\cdot\!\epsilon'^*
    + m_{22}\,\rlap/\epsilon\,\epsilon'^*\!\!\cdot\!v
    + m_{23}\,\rlap/\epsilon'^*\,\epsilon\!\cdot\!v'
    + m_{24}\,\epsilon\!\cdot\!v'\,\epsilon'^*\!\!\cdot\!v \Big] \,,
    \nonumber
\end{eqnarray}
and $m_{-+}^{M M'}$ is described by a set of related functions $m'_i$,
since $m_{-+}^{M M'} = \overline{m}_{+-}^{M' M}$. In these expressions
a ``bar'' means Dirac conjugation combined with an exchange of
velocities, polarizations, and meson masses. Also, because of radiative
corrections the functions $\ell_i$ and $m_i$ depend logarithmically on
the heavy quark masses, and $m'_i$ are related to $m_i$ by an
interchange of $m_Q$ and $m_{Q'}$ in the renormalization factors. At
tree level there is no such difference, and $\ell_i$ and $m_i$ are
universal, $m_Q$-independent functions.

The tree level expressions for these functions can be obtained by
evaluating the various traces, as explained in Sec.~\ref{sec:4}. We
find:
\begin{eqnarray}
   \ell_1 &=& (\lambda_1 + 3\lambda_2)\,\xi + B_1 + 2(w-1) B_2 + 3 B_3
    + C_1 + 2(w-1) C_2 + 3 C_3 \nonumber\\
   &&- 3 C_4 - 9 C_5 - 6 C_6 + 2(w^2-1) (2 C_7 + C_8)
    - 4(w-1) (C_9 + C_{12}) \nonumber\\
   \ell_2 &=& (\lambda_1 - \lambda_2)\,\xi + B_1 - B_3 + C_1 - C_3
    - 3 C_4 - C_5 + 2 C_6 \nonumber\\
   &&+ 2(w-1) \Big[ (w+1) C_8 - C_9 -C_{10} + 3 C_{11}
    + C_{12} \Big] \nonumber\\
   \ell_3 &=& -2 B_2 - 2 C_2 + 4(w+1) C_7 + 2 C_9 - 2 C_{10}
    - 10 C_{11} - 10 C_{12} \nonumber\\
   \ell_4 &=& - \bar\Lambda L_1 - w (\widetilde{\phi}_0
    + 3 \widetilde{\phi}_3) - (w+1)\phi_1 + 4\phi_2 + 3\phi_3
    \nonumber\\
   &&- 2 E_3 - 4(w+1) E_6 - 6 E_9 + 2(w+1) E_{10} - 4 E_{11}
    \nonumber\\
   \ell_5 &=& - \bar\Lambda L_2 - w (\widetilde{\phi}_0
    - \widetilde{\phi}_3) - (w+1)\phi_1
    + 2\phi_2 + \phi_3 + 2(w+1) E_{10} - 2 E_{11} \nonumber\\
   \ell_6 &=& - \case{2}/{w+1} \Big[ \bar\Lambda L_2
    + \case{\bar\Lambda}/{2}(w-1) L_3 + w (\widetilde{\phi}_0
     - \widetilde{\phi}_3) + (w+1)\phi_1 - \phi_2 \nonumber\\
   &&\phantom{ - \case{2}/{w+1}\, }
    - E_3 + 2(w+1) E_6 + E_9 - 2(w+1) E_{10} + E_{11} \Big] \\
&& \nonumber\\
   m_1 &=& D_1 + 2(w-1) D_2 + 3 D_3 - (2w+1) D_4 - 9 D_5 - 6 D_6
    \nonumber\\
   &&- 2(w-1) \Big[ (w+1) (2 D_7 + D_8) - 4 D_9 - 8 D_{10} \Big]
    \nonumber\\
   m_2 &=& D_1 + (w-1) D_2 + D_3 + D_4 + 3 D_5 + 2 D_6
    - 2(w-1) (D_9 + D_{10}) \nonumber\\
   m_3 &=& -2 D_2 + 2 D_4 + 2(w+1) (2 D_7 + D_8) - 2 D_9 - 10 D_{10} ,
    \nonumber\\
   m_4 &=& D_1 - D_3 + (2w-1) D_4 - D_5 - 2(2w-1) D_6 \nonumber\\
   &&+ 2(w-1) \Big[ (w+1) D_8 - 2 D_9 + 2 D_{10} \Big] \nonumber\\
   m_5 &=& -4w D_4 + 4(2w-1) D_6 -4(w-1) \Big[ (w+1) D_8 - 2 D_9
    + 2 D_{10} \Big] \nonumber\\
   m_6 &=& -2 D_2 - 2 D_4 + 4 D_6 - 2(w+1) D_8 + 6 D_9 - 2 D_{10}
    \nonumber\\
   m_7 &=& 4 D_4 - 8 D_6 - 4 D_7 + 4w D_8 - 8 D_9 + 8 D_{10} \\
&& \nonumber\\
   m_8 &=& \phi_0 + \case{6}/{w+1}\phi_3 + \case{w-1}/{w+1} \Big[
    (w+1)\phi_1 - 6\phi_2 - 2\bar\Lambda L_4 \Big] \nonumber\\
   m_9 &=& \case{1}/{w+1} \Big[ - (w+1)\phi_1 + 2(2-w)\phi_2
    + 3\phi_3 - \bar\Lambda (w-1) L_4 \Big] \nonumber\\
   m_{10} &=& \hat\phi + \case{1}/{w+1} \Big[ 3\phi_2
    + \case{3}/{2}\phi_3 + \bar\Lambda L_4 \Big] \nonumber\\
   m_{11} &=& - \phi_0 - (w+1)\phi_1 + 2\phi_2 + 2\phi_3 \nonumber\\
   m_{12} &=& 2\phi_0 + 2w\phi_1 - \case{2}/{w+1} \Big[ 2w\phi_2
    + (2w+1)\phi_3 + \bar\Lambda (w-1) L_5 \Big] \nonumber\\
   m_{13} &=& - \hat\phi + \case{1}/{w+1} \Big[ \phi_2
    + \case{1}/{2}\phi_3 + \bar\Lambda (L_4 - 2 L_5) \Big] \nonumber\\
   m_{14} &=& - \case{2w}/{w+1}\hat\phi + \case{1}/{(w+1)^2} \Big[
    4(w+1)\phi_1 - 2(3w+4)\phi_2 + (w+2)\phi_3 \nonumber\\
   &&\phantom{ -\case{2w}/{w+1}\hat\phi + \case{1}/{(w+1)^2}\, }
    + 4\bar\Lambda L_4 - 2\bar\Lambda (4-w) L_5 \Big] \\
&& \nonumber\\
   m_{15} &=& - \bar\Lambda L_1 + \widetilde{\phi}_0
    + 3 \widetilde{\phi}_3 - 2\phi_2 \nonumber\\
   &&- 2 E_3 - 4(w+1) E_6 - 6 E_9 + 2(w+1) E_{10} - 4 E_{11}
    \nonumber\\
   m_{16} &=& - \bar\Lambda L_2 + \widetilde{\phi}_0
    - \widetilde{\phi}_3 - 2 E_3 + 2 E_9 + 2 E_{11} \nonumber\\
   m_{17} &=& - \case{2}/{w+1} \Big[ \case{\bar\Lambda}/{2}(w-1) L_3
    - \phi_2 - 2(w+1) E_6 + (w+1) E_{10} + E_{11} \Big] \nonumber\\
   m_{18} &=& - \bar\Lambda L_1 + \widetilde{\phi}_0
    + 3 \widetilde{\phi}_3 - 2\phi_2 \nonumber\\
   m_{19} &=& - \case{2}/{w+1} \Big[ \bar\Lambda L_1
    - \widetilde{\phi}_0 - 3 \widetilde{\phi}_3
    + 2\phi_2 - E_3 - 2(w+1) E_6 - 3 E_9 + (w+1) E_{10} - 2 E_{11}
    \Big] \nonumber\\
   m_{20} &=& - \bar\Lambda L_2 + \widetilde{\phi}_0
    - \widetilde{\phi}_3 - 2(w+1) E_{10} + 2 E_{11} \nonumber\\
   m_{21} &=& 4(w+1) E_{10} - 4 E_{11} \nonumber\\
   m_{22} &=& - \case{1}/{w+1} \Big[ \bar\Lambda (w-1) L_3 - 2\phi_2
    - 2(w+1) E_{10} + 2 E_{11} \Big] \nonumber\\
   m_{23} &=& \case{2}/{w+1} \Big[ - \bar\Lambda L_2
    + \widetilde{\phi}_0 - \widetilde{\phi}_3
    + E_3 - E_9 - (w+1) E_{10} \Big] \nonumber\\
   m_{24} &=& - \case{4}/{w^2-1} \Big[ \case{\bar\Lambda}/{2}(w-1) L_3
    - \phi_2 + (w+1) E_6 + (w^2-1) E_{10} - w E_{11} \Big]
\end{eqnarray}
Eq.~(\ref{Erela}) ensures that there is no pole in $m_{24}$ as $w\to 1$.
Note that the first term in $\ell_1$ and $\ell_2$ compensates the mass
renormalization performed in (\ref{master}).

\appendix{Meson Form Factors}
\label{app:2}

Let us set $\varepsilon_Q=1/2m_Q$. Then to second order in the heavy
quark expansion the meson form factors $h_i$ introduced in
(\ref{formfa}) are given by:
\begin{eqnarray}
   h_+ &=& \xi + (\varepsilon_c + \varepsilon_b)\,L_1
    + (\varepsilon_c^2 + \varepsilon_b^2)\,\ell_1
    + \varepsilon_c \varepsilon_b\,(m_1-m_8) \nonumber\\
   h_- &=& (\varepsilon_c - \varepsilon_b)\,L_4
    + (\varepsilon_c^2 - \varepsilon_b^2)\,\ell_4 \\
   && \nonumber\\
   h_V &=& \xi + \varepsilon_c\,(L_2-L_5) + \varepsilon_b\,(L_1-L_4)
    \nonumber\\
   &&+ \varepsilon_c^2\,(\ell_2-\ell_5) + \varepsilon_b^2\,
    (\ell_1-\ell_4) + \varepsilon_c \varepsilon_b\,
    \big[ (m_2+m_9) - (m_{16}+m_{18}) \big] \\
   && \nonumber\\
   h_{A_1} &=& \xi + \varepsilon_c \Big(L_2-\case{w-1}/{w+1}L_5\Big)
    + \varepsilon_b\,\Big(L_1-\case{w-1}/{w+1}L_4\Big) \nonumber\\
   &&+ \varepsilon_c^2 \Big(\ell_2-\case{w-1}/{w+1}\ell_5\Big)
    + \varepsilon_b^2 \Big(\ell_1-\case{w-1}/{w+1}\ell_4\Big)
    + \varepsilon_c \varepsilon_b \Big[ (m_2+m_9)
    -\case{w-1}/{w+1}(m_{16}+m_{18}) \Big] \nonumber\\
   h_{A_2} &=& \varepsilon_c\,(L_3+L_6) + \varepsilon_c^2\,
    (\ell_3+\ell_6) + \varepsilon_c \varepsilon_b\,
    \big[ (m_3-m_{10}) - (m_{17}-m_{19}) \big] \nonumber\\
   h_{A_3} &=& \xi + \varepsilon_c\,(L_2-L_3-L_5+L_6)
    + \varepsilon_b\,(L_1-L_4) \nonumber\\
   &&+ \varepsilon_c^2\,(\ell_2-\ell_3-\ell_5+\ell_6)
    + \varepsilon_b^2\,(\ell_1-\ell_4) \nonumber\\
   &&+ \varepsilon_c \varepsilon_b\,
    \big[ (m_2+m_9) - (m_3-m_{10}) - (m_{16}+m_{18}) - (m_{17}-m_{19})
    \big] \\
   && \nonumber\\
   h_1 &=& \xi + (\varepsilon_c + \varepsilon_b)\,L_2
    + (\varepsilon_c^2 + \varepsilon_b^2)\,\ell_2
    + \varepsilon_c \varepsilon_b\,
    \big[ (m_4-m_{11}) + (m_5-m_{12}) \big] \nonumber\\
   h_2 &=& (\varepsilon_c - \varepsilon_b)\,L_5
    + (\varepsilon_c^2 - \varepsilon_b^2)\,\ell_5 \nonumber\\
   h_3 &=& \xi + \varepsilon_c\,\big[ L_2+(w-1)L_3+L_5-(w+1)L_6 \big]
    + \varepsilon_b\,(L_2-L_5) \nonumber\\
   &&+ \varepsilon_c^2\,\big[ \ell_2+(w-1)\ell_3+\ell_5-(w+1)\ell_6
    \big] + \varepsilon_b^2\,(\ell_2-\ell_5) \nonumber\\
   &&+ \varepsilon_c \varepsilon_b\,\big[ (m_4-m_{11}) + (m_5-m_{12})
    -(w-1)(m_6-m_{13})-(w+1)(m_{22}+m_{23}) \big] \nonumber\\
   h_4 &=& h_3(\varepsilon_c\leftrightarrow\varepsilon_b) \nonumber\\
   h_5 &=& \varepsilon_c\,(L_3-L_6) + \varepsilon_c^2\,(\ell_3-\ell_6)
    + \varepsilon_c \varepsilon_b\,\big[ (m_6-m_{13}) + (m_7-m_{14})
    - (m_{22}+m_{23}) \big] \nonumber\\
   h_6 &=& h_5(\varepsilon_c\leftrightarrow\varepsilon_b) \\
   && \nonumber\\
   h_7 &=& \xi + (\varepsilon_c + \varepsilon_b)\,L_2
    + (\varepsilon_c^2 + \varepsilon_b^2)\,\ell_2
    + \varepsilon_c \varepsilon_b\,(m_4-m_{11}) \nonumber\\
   h_8 &=& h_2
\end{eqnarray}
These relations are valid at tree level. The radiative corrections to
the leading and subleading terms in the heavy quark expansion have been
calculated in Refs.~\cite{Falk,FG,QCD}.

\appendix{Modified Ward Identities}
\label{app:3}

Here we derive Ward identities which relate the derivative of the
matrix elements in (\ref{Adef}) to the matrix elements in (\ref{Edef}),
in which a derivative acts on the current. These identities are needed
in Sec.~\ref{sec:4}C to express the universal functions $E'_i$ in terms
of $E_i$ and other form factors. Let us consider the following matrix
element:
\begin{equation}
   \langle M' |\,J(z)\,| M+\delta M\rangle \equiv
   \langle M' |\,J(z)\,| M \rangle + {1\over 2 m_Q}\,
   \langle M' |\,i\!\int\!{\rm d}x\,
   T\big\{\, J(z),{\cal{L}}_1(x) \,\big\} \,| M\rangle \,.
\end{equation}
$J(z)$ is a heavy quark current in the effective theory, $| M\rangle$ is
an eigenstate of ${\cal{L}}_{\rm HQET}$, and $| M+\delta M\rangle$
denotes an eigenstate of ${\cal{L}}_{\rm HQET} + \case{1}/{2 m_Q}
{\cal{L}}_1$. In contrast to (\ref{xdep}), we have
\begin{equation}
   | M+\delta M\rangle_z = \exp\Bigg[ -i\bigg(\bar\Lambda
   + {\Delta m_M^2\over 2 m_Q}\bigg) v\cdot z\Bigg]\,
   | M+\delta M\rangle_0 \,.
\end{equation}
Using this fact, we find to order $1/m_Q$
\begin{eqnarray}
   i\partial_\gamma^z\,\langle M' |\,J(z)\,| M+\delta M \rangle
   &=& i\partial_\gamma^z\,\langle M' |\,J(z)\,| M \rangle
    + {\Delta m_M^2\over 2 m_Q}\,v_\gamma\,
    \langle M' |\,J(z)\,| M \rangle \nonumber\\
   &&+ {\bar\Lambda\over 2 m_Q}\,(v-v')_\gamma\,
    \langle M' |\,i\!\int\!{\rm d}x\, T\big\{\, J(z),{\cal{L}}_1(x)
    \,\big\} \,| M\rangle \,.
\end{eqnarray}
Collecting terms of order $1/m_Q$, we thus obtain
\begin{equation}\label{equa1}
   \Big[ i\partial_\gamma^z - \bar\Lambda\,(v-v')_\gamma \Big]\,
   \langle M' |\,i\!\int\!{\rm d}x\, T\big\{\, J(z),{\cal{L}}_1(x)
   \,\big\} \,| M\rangle
   = \Delta m_M^2\,v_\gamma\,\langle M' |\,J(z)\,| M \rangle \,.
\end{equation}
On the other hand, carrying out the derivative acting on the
time-ordered product gives
\begin{eqnarray}\label{equa2}
   i\partial_\gamma^z\,\langle M' |\,i\!\int\!{\rm d}x\,
   T\big\{\, J(z),{\cal{L}}_1(x) \,\big\} \,| M\rangle
   &=& \langle M' |\,i\!\int\!{\rm d}x\, T\big\{\,
    i \partial_\gamma J(z),{\cal{L}}_1(x) \,\big\} \,| M\rangle \\
   &&- v_\gamma\,\langle M' |\, \bar h'\,\Gamma\,P_+\,
    \Big[ (i D)^2 + Z\,s_{\alpha\beta} G^{\alpha\beta} \Big]\,h\,
    | M\rangle \,. \nonumber
\end{eqnarray}
Combining (\ref{equa1}) and (\ref{equa2}), we find for the universal
form factors the relations given in (\ref{Wardid}).

\figure{\label{fig:1}
Diagrams representing the first order power corrections to meson form
factors in HQET: (a) corrections to the current, and (b) corrections to
the effective Lagrangian. The squares represent operators of order
$1/m_Q$ or $1/m_{Q'}$.}

\figure{\label{fig:2}
Diagrams representing the second order power corrections to meson form
factors in HQET: (a) corrections to the current, (b) corrections to
the effective Lagrangian, and (c) mixed corrections to the current and
the effective Lagrangian. The black squares represent operators of order
$1/m_Q$ or $1/m_{Q'}$, the open ones denote operators of order
$1/m_Q^2$, $1/m_{Q'}^2$, or $1/m_Q m_{Q'}$.}

\end{document}